\documentclass[alpha-refs]{nbdt-article}

\papertype{Original Article}
\paperfield{Journal Section}
\usepackage{amsfonts}
\usepackage{hyperref}

\newcommand{\e}{{\rm e}}

\renewcommand{\d}{{\rm d}}
\newcommand{\pd}{\partial}
\newcommand{\PP}{{\rm P}}

\newcommand{\D}{\displaystyle}

\newcommand{\mc}{\mathcal }

\newcommand{\sign}{{\rm sign}}

\renewcommand{\S}{{\mc S}}

\newcommand{\T}{{\mc T}}
\newcommand{\lcli}[1]{t_{#1}^L}              
\newcommand{\rcli}[1]{t_{#1}^R}                 

\definecolor{myblue2}{rgb}{.0313, .988, 0.941}

\definecolor{mygreen}{RGB}{28,172,0} 
\definecolor{mylilas}{RGB}{170,55,241}

\newcommand{\hzd}{h}  	
\newcommand{\disclin}{\gamma}
\newcommand{\PC}[2]{{\rm Pr}\left(#1 \ \big | \ #2\right)} 
\newcommand{\LE}{\left}
\newcommand{\RI}{\right}
\newcommand{\train}{\mathcal{T}}                
\newcommand{\Data}{\mathfrak{D}}

\newcommand{\ldl}{\lambda_\text{low}}
\newcommand{\ldh}{\lambda_\text{high}}

\newcommand{\SNR}{{\rm SNR}}
\newcommand{\E}{{\rm E}}
\newcommand{\Var}{{\rm Var}}

\definecolor{darkgreen}{rgb}{0,0.6,0}
\definecolor{violet}{rgb}{0.8,0,0.8}
\definecolor{blueviolet}{rgb}{.6,.22,.9}

\title{Performance of normative and approximate evidence accumulation on the dynamic clicks task}

\author[1,\authfn{1}]{Adrian E. Radillo}
\author[2,\authfn{1}]{Alan Veliz-Cuba}
\author[3,4,\authfn{2}]{\\ Kre\v{s}imir Josi\'{c}}
\author[5,6,\authfn{2}]{Zachary P. Kilpatrick}

\affil[1]{Department of Neuroscience, University of Pennsylvania, Philadelphia, PA 19104}
\affil[2]{Department of Mathematics, University of Dayton, Dayton, OH 45469}
\affil[3]{Departments of Mathematics and Biology and Biochemistry, University of Houston, Houston, TX 77204}
\affil[4]{Department of BioSciences, Rice University, Houston, TX 77251, USA}
\affil[5]{Department of Applied Mathematics, University of Colorado, Boulder, CO 80309}
\affil[6]{Department of Physiology and Biophysics, University of Colorado School of Medicine, Aurora, CO 80045}
\contrib[\authfn{1}]{co-first authors}
\contrib[\authfn{2}]{co-last authors \\ {\bf Code Availability:} Codes developed to produce figures are available at \url{https://github.com/aernesto/NBDT_dynamic_clicks} \\ \vspace{-7mm}}

\corraddress{Zachary P. Kilpatrick, Department of Applied Mathematics, University of Colorado, Boulder, CO 80309}
\corremail{zpkilpat@colorado.edu}

\fundinginfo{This work was supported by NSF/NIH CRCNS (R01MH115557) and NSF (DMS-1517629). ZPK was also supported by NSF (DMS-1615737). KJ was also supported by NSF (DBI-1707400). AVC was supported by the Simons Foundation (516088) and OSC (PNS0445-2). }

\runningauthor{Radillo, Veliz-Cuba, Josi\'{c}, \& Kilpatrick}

\begin{document}

\maketitle


\begin{abstract}
The aim of a number of psychophysics tasks is to uncover how 
mammals make decisions in a world that is in flux.  Here we examine the characteristics of ideal and near--ideal  observers in a task of this type. We ask when and how performance depends on task parameters and design, and, in turn, what observer performance tells us about their decision-making process.  In the dynamic clicks task subjects hear two streams (left and right) of Poisson clicks with different rates.  Subjects are rewarded when they correctly identify the side with the higher rate, as this side switches unpredictably.
We show that a reduced set of task parameters defines regions in parameter space in which optimal, but not near-optimal observers, maintain constant response accuracy.
We also show that for a range of task parameters an approximate normative model must be finely tuned to
reach near-optimal performance, illustrating a potential way to distinguish between normative models and their approximations. 
In addition, we show that using the negative log-likelihood and the 0/1-loss functions to fit these types of models is not equivalent: the 0/1-loss leads to a bias in parameter recovery that increases with sensory noise. 
These findings suggest  ways to tease apart models that are hard to 
distinguish when tuned exactly, and point to general pitfalls in experimental design, model fitting, and interpretation of the resulting data. 

\end{abstract}


\section{Introduction}

Decision-making tasks are widely used to probe the neural computations that underlie behavior and cognition~\citep{Luce86,gold07}.
Mathematical models of optimal decision-making (normative models)\footnote{We will use the phrases `optimal model,' `optimal observer,' `normative model,' and `ideal observer' interchangeably, as they refer to the best possible model for a given set of task and observation constraints.} have been key in helping us understand tasks that require the accumulation of noisy evidence~\citep{wald48,gold02,Bogacz2006}.
 Such models assume that an observer integrates a sequence of noisy measurements to determine the probability that one of several options is correct~\citep{wald48,beck08,velizcuba16}.

The random dot motion discrimination (RDMD) task is a prominent example, in which the neural substrates of the evidence accumulation process can be identified in cortical recordings~\citep{ball82,britten92,Roitman2002}.  The associated normative models take the form of tractable stochastic differential equations~\citep{ratcliff78,Bogacz2006}, and have been used to explain behavioral data~\citep{ratcliff08,krajbich11}. Neural correlates of subjects' decision processes display striking 
similarities with these models~~\citep{shadlen96,huk05}, although a clear link between the two is still under debate~\citep{latimer15,shadlen16}.

Poisson clicks tasks~\citep{Brunton2013,Odoemene2017} have recently become popular in studying the cortical computations underpinning mammalian perceptual decision-making. Neural activity during such tasks also appears to reflect an underlying evidence accumulation process~\citep{Hanks2015}. The corresponding normative models  and their approximations are low-dimensional  and computationally tractable. This makes the task well-suited to the analysis of data in high-throughput experiments~\citep{Brunton2013}. 
\cite{piet18} have extended the clicks task to a dynamic environment to understand how animals adjust their  evidence accumulation strategies when older evidence decreases in relevance. \cite{glaze15} carried out a similar study in an extension of the RDMD task. Both studies concluded that subjects are capable of implementing evidence accumulation strategies that adapt to the timescale of the environment.

However, identifying the specific strategy subjects use to solve a decision task can be difficult because different strategies can lead to similar observed outcomes~\citep{ratcliff08}. How to set task parameters to best identify a subject's evidence accumulation strategy has not been studied systematically, especially in dynamic environments~\citep{ratcliff16}. Here, we focus on the dynamic clicks task and aim to understand what task parameters (or combinations of parameters) determine performance, and under what conditions different strategies can be identified.

In the dynamic clicks task, two streams of auditory clicks are presented simultaneously to a subject, one stream per ear~\citep{piet18,Brunton2013}. Each click train is generated by a Markov-modulated Poisson process~\citep{Fischer1992} whose click arrival rates switch between two possible values ($\lambda_\text{low}$ vs. $\lambda_\text{high}$). The two streams have distinct rates which switch at discrete points in time according to a memoryless process with hazard rate $h$. Thus streams played in different ears always have distinct  rates ($\lambda^L(t) \neq \lambda^R(t)$). The subject must choose the stream with highest instantaneous rate when interrogated at a time $T$, which ends the trial. Switches occur at random times that are not signaled to the subjects, who must therefore base their decision on the observed sequences of Poisson clicks alone. 
The rate $h$ at which the environment changes is a latent variable that needs to be learned for optimal performance.
In this study, however, our observer models always use a constant rate of change for their environment
\footnote{See \cite{Radillo2017} for an optimal observer that can learn the hazard rate $h$ in a dynamic version of the RDMD task. This approach can be extended to the case of the dynamic clicks tasks as in \cite{radillo17diss}.}. 

We analyze the normative model of the dynamic clicks task to shed light on how its response accuracy depends on task parameters, as this is a measure commonly used when fitting to behavioral data~\citep{piet18}.
As shown in Section \ref{sec:norm}, the ideal observer accumulates evidence from each click to update their log likelihood ratio (LLR) of the two choices. Each click corresponds to a pulsatile increase  or decrease in the LLR. Importantly, evidence must be discounted at a rate that accounts for the timescale of environmental changes.

The main goal of this work is to identify how task parameters shape an ideal observer's response accuracy, and the identifiability of evidence accumulation models.
We find effective parameters that can be fixed to keep the accuracy of the ideal observer constant.\footnote{We use the term ``accuracy'' to refer to the percentage of correct choices for a given model and parameter set. This is our primary measure of a model's performance on the task.}
One such parameter is the signal-to-noise ratio (SNR) of the clicks during a single epoch between changes and the other is $hT$, the trial length $T$ rescaled by the hazard rate $h$. These two parameters fully determine the accuracy of an  optimal observer interrogated at the end of the trial (Section \ref{sec:SNR}), as well as response accuracy conditioned on the time since the final change point of a trial (Section \ref{sec:change}). 

While the normative model determines the optimal strategy, subjects may also use heuristics or approximations that are potentially simpler to implement. The accuracy of approximate models may also be more sensitive to parameter changes, so fitting procedures converge more rapidly. As an example we consider a linear model, which has been previously fit to data from subjects performing dynamic decision tasks~\citep{piet18,glaze15}, and has also been studied as an approximation to normative evidence accumulation~\citep{velizcuba16}. To obtain response accuracy close to that of the normative model, the discounting rate of the linear model needs to be tuned for different click rates and hazard rates (Section~\ref{sec:linear}). In contrast, the discounting rate in the normative (nonlinear) model equals the hazard rate. Moreover, the linear model's accuracy is more sensitive to changes in its evidence-discounting parameter than the nonlinear model.
\footnote{The `nonlinear' model here refers to the family of models obtained by tuning the discounting rate away from the value defining the normative model. This detuning results in a model that is not normative.}
This effect is most pronounced at intermediate SNR values, suggesting a task parameter range where the two models can be distinguished.

Lastly,
we ask how model parameters can be inferred from subject responses. Using maximum likelihood fits of the models to choice data, we show that the fit discounting parameters are  closer to the true parameter in the linear model compared to the nonlinear model
(Section~\ref{sec:identify}). This is expected, since the response accuracy of the nonlinear model depends weakly on its discounting parameter. 
We also explore the impact of the loss function on model fitting, and  show that in the presence of sensory noise using a 0/1-loss function  results in a systematic bias in parameter recovery (Section~\ref{S:fitting}).  
The 0/1-loss function gives a one unit penalty on trials in which the decision predicted by the model and the data disagree, and no penalty when they agree. Therefore, minimizing this loss function leads to models that best match the trial-to-trial responses in the data rather than the response accuracy.

Ultimately, our findings point to ways of identifying task parameters for which subjects' decision accuracy is sensitive to the mode of evidence accumulation they use in fluctuating environments.  We also show how using different models and different data fitting methods can lead to divergent results, especially in the presence of sensory noise. We argue that similar issues can arise whenever we try to interpret data
from decision-making tasks.


\section{Normative model for the dynamic clicks task}
\label{sec:norm}

\begin{figure}[!t]
\centering
\includegraphics[scale=0.35]{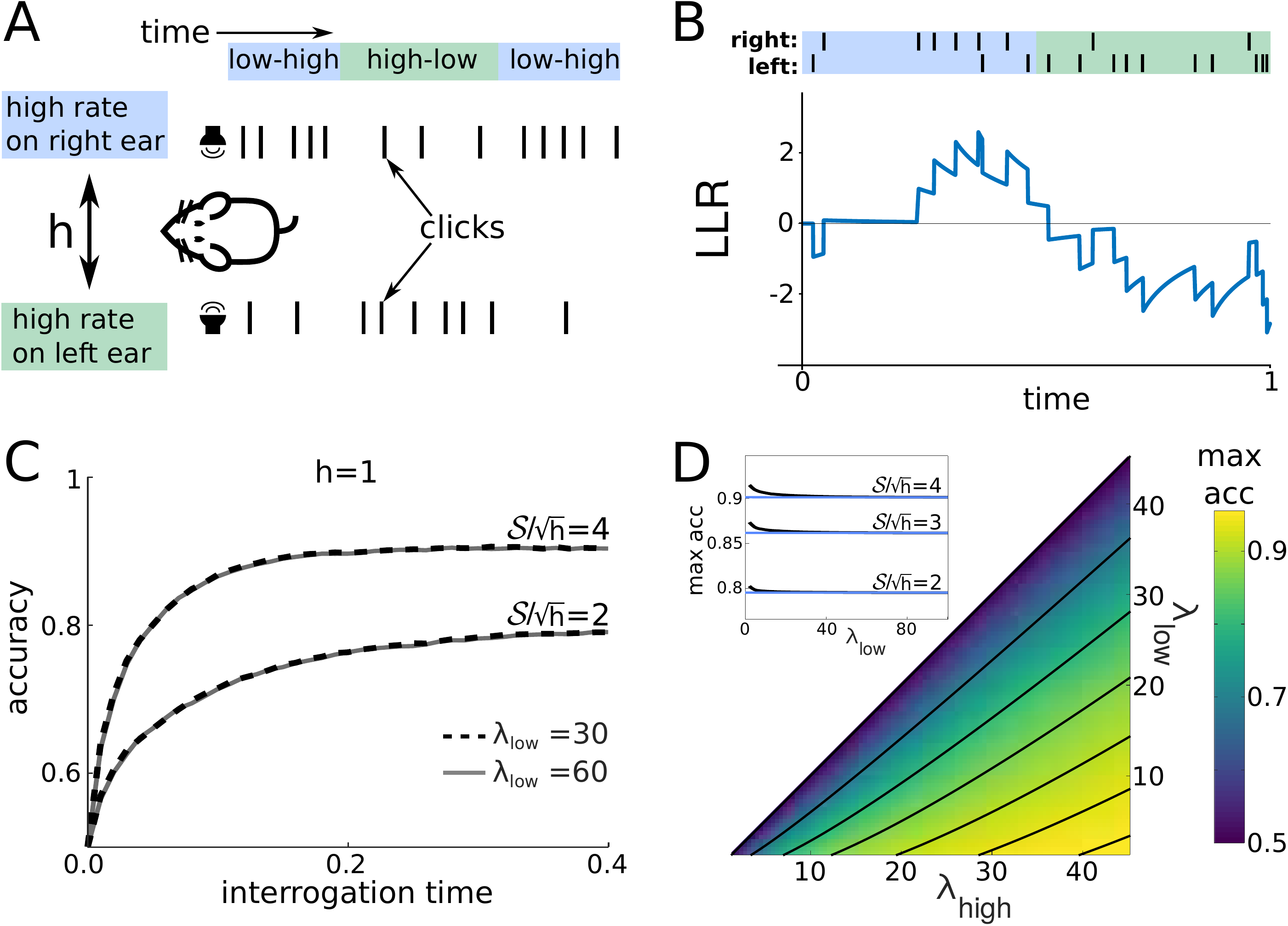}
\caption{ 
{\bf A:} Schematic of the dynamic clicks task from \cite{piet18}. 
{\bf B:} A single trajectory of the log-likelihood ratio (LLR), $y_t,$ during a trial. The click streams and environment state are shown above the trajectory. 
{\bf C:} Response accuracy of the ideal observer  as a function of interrogation time for two distinct SNR values, $\S/\sqrt{h},$  defined in Eq.~\eqref{fixparams}. Two distinct pairs of click rates used in simulations ($\ldl=30$ and $60$ Hz)
were chosen to match each SNR at hazard rate $h = 1$ Hz, resulting in overlaying dashed ($\ldl=30$Hz) and solid  ($\ldl=60$Hz) lines. For $\S/\sqrt{h}=2$, we take $(\ldh, \ldl)= (58.17, 30)$ and $(97.67,60)$; for $\S/\sqrt{h}=4$, $(\ldh, \ldl)= (70, 30)$ and $(112.54,60)$. Fixing $c_1$ and $c_2$ in Eq.~(\ref{fixparams}) yields a match in accuracy for both pairs of click rates. As time evolves during the trial, accuracy saturates. Note $hT$ and ${\mc S}/\sqrt{h}$ jointly determine accuracy.
{\bf D:} Maximal accuracy of the ideal observer at $T \gg 1$ is constant along level curves (black lines) of $\S / \sqrt{h}$ (seen as a function of $(\ldl,\ldh)$ with constant $h$) across a wide range of parameters, consistent with Eq.~\eqref{fixparams}. We only show 
the $(\ldl, \ldh)$ region
where $0<\ldl<\ldh$. Level curves of  $\S$ are oblique parabolas in the $(\ldl,\ldh)$ plane. {\bf D(Inset):} Level curves slightly underestimate accuracy for small $\ldh$ and $\ldl$ (black: maximal accuracy; blue: maximal accuracy for large values of $\ldh$ and $\ldl$). See Appendix \ref{app:mc} for details on Monte Carlo simulations.}
\label{fig1_snr}
\end{figure}

In the dynamic clicks task an observer is presented with two Poisson click streams, $s^L(t)$ and $s^R(t)$ ($0<t\leq T$), and needs to decide which of the two has a higher rate~\citep{Brunton2013}.  The rates of the two streams are not constant, but change according to a hidden, continuous-time Markov chain, $x(t)$, with binary state space $\{x^R,x^L\}$. The frequency of the switches is determined by the \emph{hazard rate}, $h$, so that $\PP(x(t+dt) \neq x(t)) = h \cdot dt + o(dt)$.
The left and right rates, $\lambda^L(t)$ and $\lambda^R(t)$, 
can each take on one of two values, $\{ \lambda_{\rm high}, \lambda_{\rm low}\}$ with $\lambda_{\rm high} > \lambda_{\rm low}>0$. 
When $x(t)=x^L$, we have $(\lambda^L(t),\lambda^R(t))=(\lambda_\text{high},\lambda_\text{low})$, and when $x(t)=x^R$ the opposite is true. Therefore $x(t) = x^k$ means that stream $k$ has the higher rate at time $t$: $\lambda^k(t) = \lambda_{\rm high}$  (Fig.~\ref{fig1_snr}A). The observer is prompted to identify the side of the higher rate stream, $x(T)$, at a random time $T$.  
The interrogation time, $T$, is sampled ahead of time by the experimenter for each trial and is unknown to the subject. We refer the reader to \cite{piet18} and \cite{Brunton2013} for more details about the experimental setup.

This task is closely related to the \emph{filtering of a Hidden Markov Model} studied in the signal processing literature~\citep{Cappe2005, rabiner86}. 
For a single, 2-state Markov-modulated Poisson process~\citep{Fischer1992},  the filtering problem was solved by \cite{rudemo72} -- see also \citep{snyder75} for review and extensions.
This filtering problem corresponds to a task in which a single, variable rate click stream is presented to the observer who has to report whether at some time $T$ the rate is high or low. In the present case, the observer is presented with two coupled Markov-modulated Poisson processes. The normative model reduces to that considered by \cite{rudemo72} when we consider a single stream version of the task, so our approach can be considered a generalization.

Assuming the Poisson rates $\{ \lambda_{\rm high}, \lambda_{\rm low} \},$ and the hazard rate, $h,$ are known, a normative model for the inference of the hidden state, $x(t)$, has been  derived by~\cite{piet18}. The resulting model
can be expressed as an ordinary differential equation (ODE) describing the evolution of the LLR of the two environmental states: 
\begin{equation}
y_t:=\log\frac{\PP(x(t)=x^R|s^R(t),s^L(t))}{\PP(x(t)=x^L|s^R(t), s^L(t))}. \label{eq:LPOR}
\end{equation}
For completeness, we present the derivation in Appendix~\ref{app:changing}, yielding the same ODE as \cite{piet18}:
\begin{equation}
\frac{dy_t}{dt}=\kappa \underbrace{ \left[ \sum_{i=1}^{\infty} \delta(t-t^R_i)- \sum_{j=1}^{\infty} \delta(t-t^L_j) \right]}_\text{right and left click streams}  - \underbrace{2h\sinh(y_t)}_\text{nonlinear discounting},
\label{eq:ode_changing}
\end{equation}
where $\kappa:=\log(\lambda_\text{high}/\lambda_\text{low})$ is the evidence gained from a click, $\delta(t)$ is the Dirac delta function centered at 0, and $t^R_i$ (resp. $t^L_j$) is the $i$-th right click (resp. $j$-th left click).

Eq.~(\ref{eq:ode_changing}) has an intuitive interpretation: A click provides evidence that the higher rate stream is on the side at which the click was heard. Thus, a click heard on the right (left) results in a positive (negative) increment in the LLR (Fig.~\ref{fig1_snr}B). Since the environment is volatile,
as evidence recedes into the past it becomes less relevant. In Eq.~(\ref{eq:ode_changing}) each click is followed by a  superlinear decay to zero. Note that  the discounting term only depends on the current LLR, $y_t$, and the hazard rate, $h$, and not on the click rates. 

Performance on the task may increase with the informativeness of each click, $\kappa = \log (\lambda_{\rm high}/\lambda_{\rm low})$. However, $\kappa$ alone does not predict the response accuracy (i.e. the fraction of correct trials) of the normative model~\citep{Brunton2013,piet18}. In the next section, we will show that an ideal observer's response accuracy is determined by the click frequencies $(\ldh, \ldl)$ and the hazard rate $h$:  A sequence of a few very informative clicks may provide as much evidence as many clicks, each carrying little information. But if the environmental hazard rate is high, even informative clicks quickly lose their relevance.

The LLR, $y_t$, contains all the information an ideal observer has about the present state of the environment, given the observations~\citep{gold02}. If interrogated at time $t=T$,  $\sign (y_T) = \pm 1$, determines the most likely current state ($x^R$ for $+1$ and $x^L$ for $-1$), and therefore the response of an optimal decision maker.  In the following, we will show that  two effective parameters govern the response accuracy of the optimal observer.
\vfill
\pagebreak


\section{The signal-to-noise ratio of dynamic clicks}
\label{sec:SNR}

Four parameters characterize the dynamic clicks task: the hazard rate, $h$, duration of a trial, \emph{i.e.} interrogation time, $T$, the low click rate, $\ldl$, and the high click rate, $\ldh$. 
However, we next show that only two effective parameters typically govern an ideal observer's performance  (Fig.~\ref{fig1_snr}C,D): the product of the interrogation time and the hazard rate, $hT$, and  the signal-to-noise ratio (SNR) of the dynamic stimulus. The former corresponds to the mean number of switches in a trial, and the latter combines the click rates $\ldl$ and $\ldh$ into
a Skellam--type SNR (Eq.~(\ref{skellam}) below), scaled by the hazard rate $h$ (Eq.~\eqref{fixparams}). 

To motivate our definition, consider first the case of a static environment, $h=0$ Hz, for which the normative model is given by Eq.~\eqref{eq:ode_changing} without the nonlinear term. Since $\kappa$ does not affect the sign of $y_t$, response accuracy depends entirely on the difference in click counts $N^R(t) - N^L(t)$, where $N^j(t)$ are the counting processes associated with each click stream. Thus we can define the difference in click counts as the \emph{signal}, and the SNR as the ratio between the signal mean and standard deviation at time $T$~\citep{skellam46},
\begin{align}
\text{SNR}_0^T : = \frac{\E[N^R(T)] - \E [ N^L(T)]}{\sqrt{\Var[ N^R(T)] + \Var [N^L(T)]  }} =  \frac{T\lambda_\text{high}-T\lambda_\text{low}}{\sqrt{T\lambda_\text{high}+T\lambda_\text{low}}} =  \S \cdot \sqrt{T},  \label{SNRh0}
\end{align}
where 
\begin{align}
{\mc S}:= \frac{\lambda_\text{high}-\lambda_\text{low}}{\sqrt{\lambda_\text{high}+\lambda_\text{low}}}. \label{skellam}
\end{align}

In a dynamic environment, the volatility of the environment, governed by the hazard rate, $h$, also affects response accuracy.
The environment can switch states immediately before the interrogation time, $T$, decreasing response accuracy. This suggests that accuracy will not only be determined by the click rates, but also by the length of time the environment remains in the same state prior to interrogation. Using this fact and the definition of SNR  in a static environment, we determine the statistics for the difference in the number of clicks between the high- and low-rate streams during 
the final epoch preceding interrogation (for derivation details see Appendix \ref{SNR_derive}). Averaging over the Poisson distributions characterizing the click numbers, and the epoch length distribution yields a nonlinear expression  representing the SNR that involves $\S/ \sqrt{h},$ and the rescaled trial time $hT$:
\begin{align}
F(hT, \S/ \sqrt{h}) = \frac{(1 - \e^{-hT}) \S/ \sqrt{h}}{\sqrt{ (1 - 2 hT \e^{-hT} - \e^{-2 hT}) \S^2/h + 1 - \e^{-hT} }}.
\label{E:SNRh}
\end{align}

The unitless measure of trial duration, $hT,$ characterizes the timescale of the evolution of the LLR, $y_t$.  As accuracy  should not depend on the units in which we measure time, this is a natural measure of the evidence accumulation period~\footnote{This is related to dimensional analysis often used when studying physical models~\citep{langhaar80}.}.  
As indicated, $F(hT, {\mc S}/\sqrt{h})$ only depends on $hT$ and ${\mc S}/ \sqrt{h}$.
We therefore predict that optimal observer response accuracy is determined by the following 
 two parameter combinations, 
\begin{align}
\S/\sqrt{h} =  \frac{\lambda_{\rm high} - \lambda_{\rm low}}{\sqrt{h(\lambda_{\rm high} + \lambda_{\rm low})}} = c_1 \hspace{6mm} \text{and} \hspace{6mm} hT = c_2.  \label{fixparams}
\end{align}
Henceforth, we will refer to ${\mc S}/\sqrt{h}$ as the SNR and $hT$ as rescaled trial time. Note that the term ${\mc S}/\sqrt{h}$ can also be realized as a SNR of Eq.~(\ref{eq:ode_changing}) by performing a diffusion approximation, and computing the SNR of the corresponding drift-diffusion signal (See Appendix~\ref{app:diffapprox}). 

 Fig.~\ref{fig1_snr}C shows examples in which the ideal observer's response accuracy is constant when SNR and $hT$  are fixed.
Accuracy is computed as the fraction of trials at which the observer's belief, $y_t,$ matches the underlying state, $x(t),$ at the interrogation time, $T$, that is the fraction of  trials for which ${\rm sign}(y_T) = x(T)$.
  The accuracy as a function of $T$  and $h = 1$ remains constant if we change $\ldh$ and $\ldl$, but keep $\S$ fixed.
As the interrogation time $T$ is increased, the accuracy saturates to a value below 1 (Fig.~\ref{fig1_snr}C), consistent with previous modeling studies of decision-making in dynamic environments~\citep{glaze15,velizcuba16,Radillo2017,piet18}.
Evidence discounting limits the magnitude of the LLR, $y_t$. Hence a sequence of low rate clicks can lead to errors, especially for low SNR values. Moreover, on some trials the state, $x(t),$ switches close to the interrogation time $T$.  As it may take multiple  clicks for $y_t$ to change sign after a change point (See Fig.~\ref{fig1_snr}B), this can also lead to an incorrect response.

In Fig.~\ref{fig1_snr}D we show that the maximal accuracy (obtained for $T$ sufficiently large) as a function of $\ldh$ and $\ldl$ (colormap), is approximately constant along SNR level sets (black oblique curves). This correspondence is not exact when $\ldh$ and $\ldl$ are small (Fig.~\ref{fig1_snr}D inset), and we conjecture that this is because higher order statistics of the signal determine response accuracy in this parameter range. As discussed in Appendix \ref{app:diffapprox}, for large $\ldh$ and $\ldl$ we can use a diffiusion approximation for the dynamics of Eq.~(\ref{eq:ode_changing}). When $\ldh$ and $\ldl$ are small, the diffusion approximation does not apply, and response accuracy is characterized by features of the signal beyond its mean and variance. Since the SNR only describes the ratio between the mean and standard deviation of the stimulus, it cannot capture the impact of higher order statistics on accuracy at low click rates. Nonetheless, the SNR predicts response accuracy well.

The consequences of these observations are twofold: Two parameter combinations
determine optimal performance, potentially simplifying experiment design. 
To ensure coverage of different response accuracy regimes, we can initially vary SNR and $hT$.
To increase the accuracy of an ideal observer, it is not sufficient to increase both click rates, for instance, since the SNR stays constant if $\lambda_{\rm high}$ and $\lambda_{\rm low}$ follow  the parabolas shown in Fig.~\ref{fig1_snr}D.
Second, this approach makes testable predictions about the
accuracy of an optimal observer:  If we change parameters so that SNR and $hT$ are fixed, and a subject's accuracy is affected, this indicates that the subject may not have learned the hazard rate, $h$ or is using a suboptimal discounting model.


\section{Post change-point decisions depend on SNR}
\label{sec:change}

\begin{figure}[!t]
\centering
\includegraphics[scale=0.3]{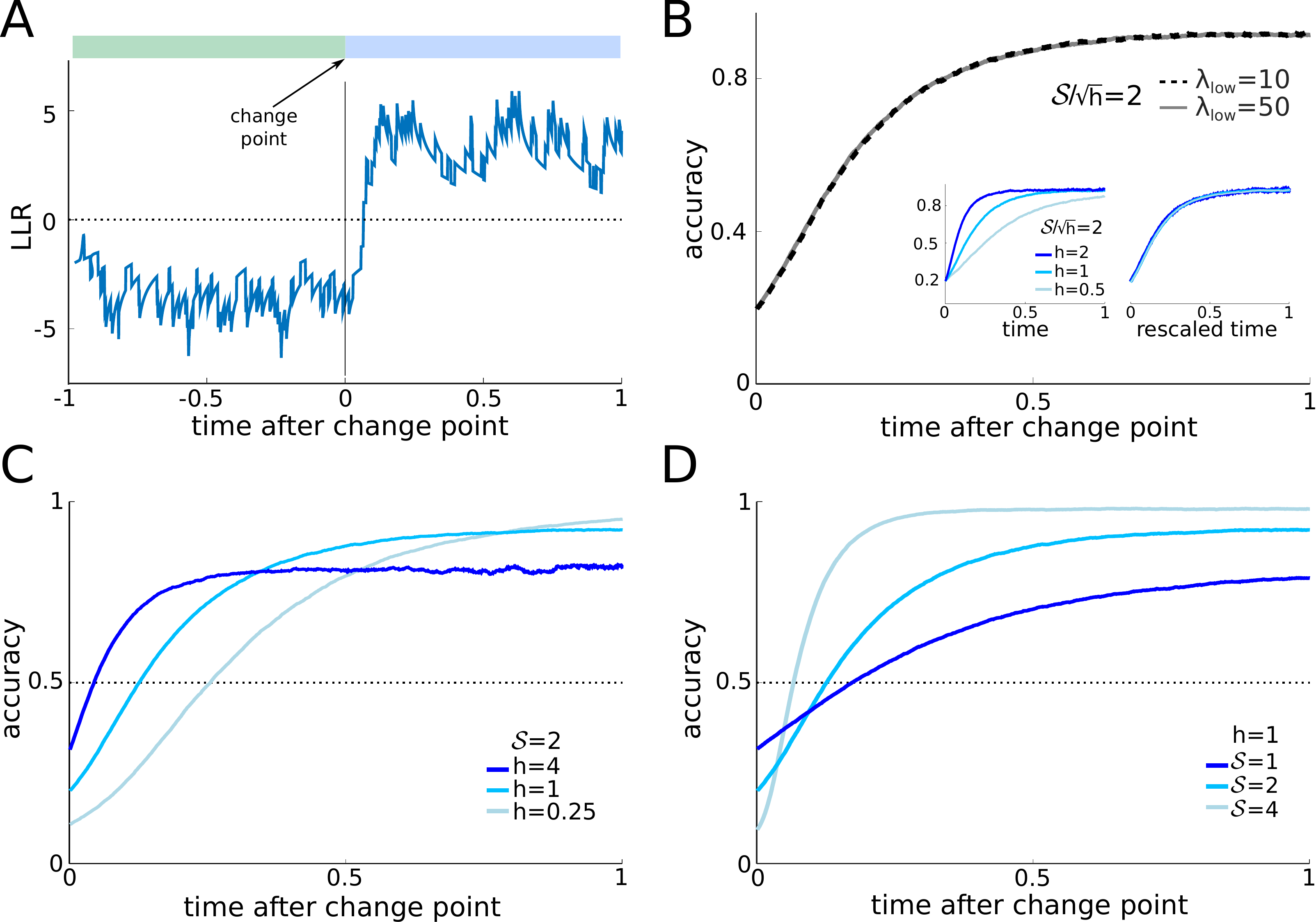}
\caption{
{\bf A:} After a change point, evidence in favor of the correct hypothesis has to be accumulated before optimal observers change their belief. Thus, there is a delay between the change point and the sign change of the LLR.
{\bf B:}~The  accuracy as a function of time following a change point is the same when $\S / \sqrt{h}$ and $h$ are fixed ($h=1$ in the plot). 
{\bf B(inset):}
Furthermore  accuracy is constant for fixed $\S / \sqrt{h}$ and rescaled response time $h T_{\rm fin}$  (left: time without rescaling, right: time after rescaling).
{\bf C:} For lower hazard rates, $h$, accuracy at  change point time is lower, and requires a longer time to reach 0.5.
{\bf D:} As the strength of evidence,  $\S$, increases,  accuracy at the time of the change point is lower, but increases more rapidly.
In panels C and D the value of $\S/ \sqrt{h}$ is the same for curves of the same color. Rescaling the timescale of curves in panel C by $h$ would yield the curves in panel D. 
}
\label{fig2_change}
\end{figure}

To understand how an optimal observer adapts to environmental changes, we next ask how their fraction of correct responses depends on the final time, $T_{\rm fin},$ between the last change point preceding a decision and the decision itself (Fig. \ref{fig2_change}A). Overall accuracy again depends on both SNR and rescaled trial time $hT$. In addition, for sufficiently long trials, accuracy as a function of time since the last change point depends only on the rescaled time since the change point, $hT_{\rm fin}$ and the SNR.

 If the click rates, $\ldh$ and $\ldl,$ are varied, but $\S$ and $h$ are held fixed, the accuracy as a function of $T_{\rm fin}$  remains unchanged (Fig.~\ref{fig2_change}B, for $h=1$, $\S=2$).  On the other hand, accuracy changes if we fix $\S/\sqrt{h}$ (SNR) but vary $h$  (Fig.~\ref{fig2_change}B, left inset). With fixed $\S/\sqrt{h},$  accuracy depends only on the rescaled time since the last change point, $hT_{\rm fin}$  (Fig.~\ref{fig2_change}B, right inset). Thus, while absolute accuracy depends on the total length of the trial, $T,$ measured in units of average epoch length, $1/h$, accuracy relative to the last change point depends only on the elapsed time, $T_{\rm fin},$ measured in the same units.

\cite{glaze15} introduced the notion of an accuracy {\em crossover effect} in the dynamic RDMD task: The normative model predicts that after a change point observers update their belief more slowly, but eventually reach higher accuracy at low compared to high hazard rates. Thus plotting the maximal accuracy against time since the last change point for different hazard rates results in curves that cross. Behavioral data indicates that human observers behave according to this prediction~\citep{glaze15,glaze18}.

A similar crossover effect also occurs in the dynamic clicks task: Accuracy just after a change point is lower for small hazard rates, $h,$ (Fig. \ref{fig2_change}C) and takes longer to reach 50\%, but saturates at a higher level compared to more volatile environments. In slow environments, the optimal observer integrates evidence over a longer timescale $(1/h)$, leading to more reliable estimates of the state, $x(t)$. But this increased certainty comes at a price, as  it requires more time to change the observer's belief after a change point.   Similarly, in environments with stronger evidence (larger $\S$, Eq.~(\ref{skellam})), accuracy immediately following a change point is lower, since state estimates, and hence the beliefs are more reliable compared to trials with weak evidence (Fig. \ref{fig2_change}D). However, stronger evidence also  causes a rapid increase in accuracy, which then saturates at a higher level than on trials with weaker evidence (lower $\S$). Therefore, both evidence quality, and environmental volatility determine accuracy after a change point.

We conclude that accuracy after a change point is characterized by SNR ($\S/ \sqrt{h}$) and the rescaled time since the change point, $hT_{\rm fin}$. This only holds when trials are sufficiently long, and the belief at trial outset does not affect accuracy.  In addition,  increasing SNR lowers accuracy immediately after a change point, and increases the recovery of accuracy to a higher saturation point (Fig. \ref{fig2_change}D). On the other hand decreasing volatility, $h,$ while fixing ${\mc S}$ (Fig. \ref{fig2_change}C) leads to lower accuracy immediately after a change point, and higher saturation.  However, the rate at which accuracy is recovered after a change point \emph{decreases} with decreasing $h$. 

These are again characteristics of an optimal observer, and  deviations from these predictions  indicate departures from optimality.


\section{A linear approximation of the normative model}
\label{sec:linear}

\begin{figure}[!t]
	\centering
	\includegraphics[scale=0.37]{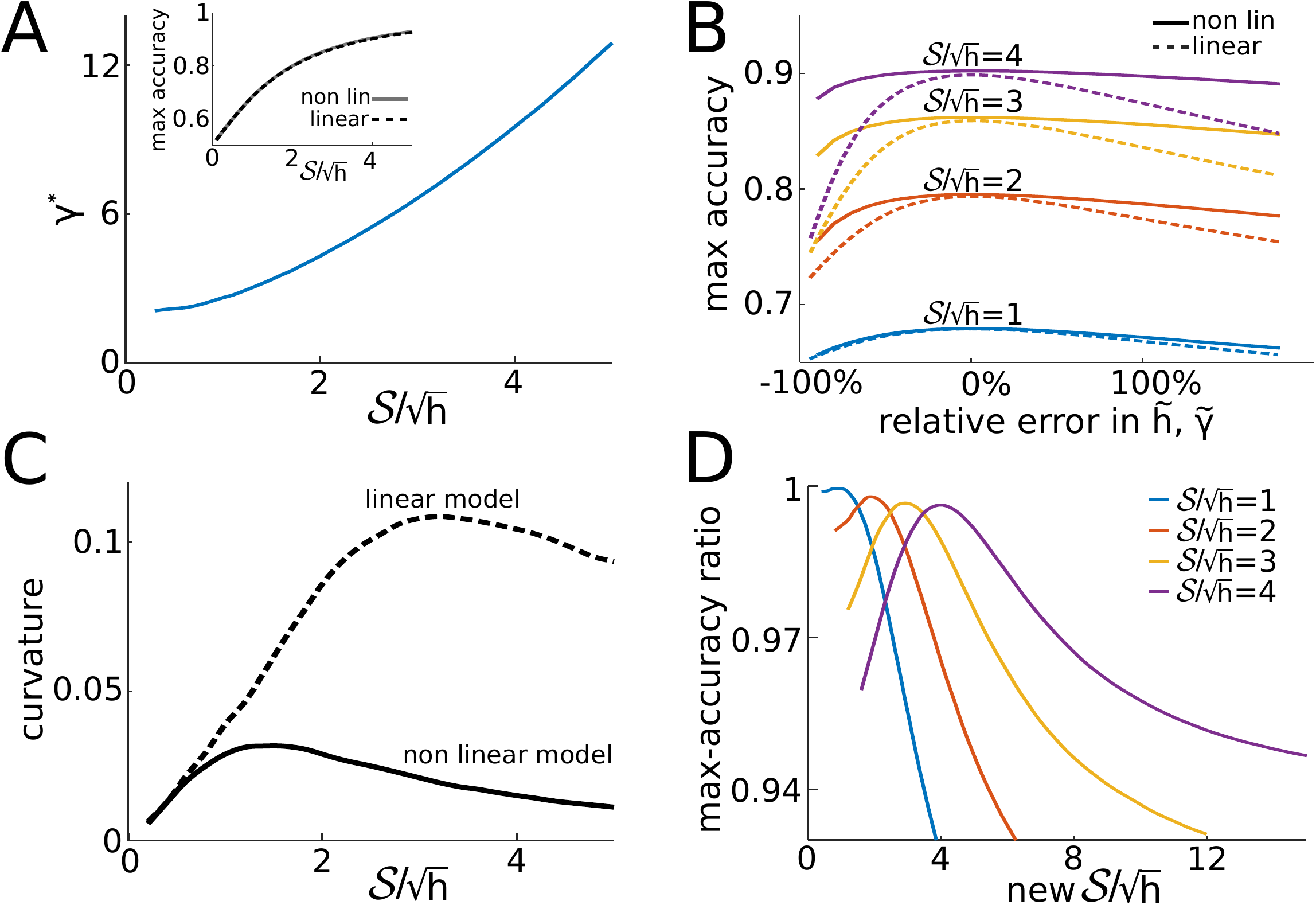}
	\caption{
		{\bf A:} Optimal linear discounting rate, $\gamma^*$ in Eq.~\eqref{eq:ode_linear}, as a function of $\S/ \sqrt{h}$. 
		{\bf A(inset):}
		The accuracy of the linear and nonlinear model are nearly identical over a wide range of SNR values, $\S/ \sqrt{h},$ when the linear discounting rate is set to  $\gamma^*$ (here $h=1$ is fixed). 
		{\bf B:} Response accuracy near the optimal discounting rate for the linear model (dashed), and assumed hazard rates for the nonlinear model (solid) for several SNR values ($h=1$). The linear model is  more sensitive to relative changes in the discounting rate. The relative error is defined as $100(\tilde{h}-h)/h$ for the nonlinear model and $100(\tilde{\gamma}-\gamma^*)/\gamma^*$ for the linear model.
		{\bf C:} Curvature (absolute value of the second derivative) of the accuracy profiles in panel {\bf B}, evaluated at their peak, as a function of $\S / \sqrt{h}$. The curvature, and hence sensitivity, of the nonlinear model is higher for intermediate and large values of $\S/ \sqrt{h}$.
		Since the functions in panel \textbf{B} do not depend on the actual values of $\tilde{h}$ and $\tilde{\gamma}$, but rather the relative distance of these parameters from reference values, what we show in this plot are \emph{relative} curvatures. We compare relative curvatures as $\tilde{h}$ and $\tilde{\gamma}$ do not have the same units.
		{\bf D:} Ratio of the accuracy of the linear model to that of the normative model, as SNR is varied. Along each curve, the discounting rate $\tilde{\gamma}$ of the linear model is held fixed at the value $\gamma^*$ that would maximize accuracy at the reference SNR indicated by the legend.
	}
	\label{fig3_linear}
\end{figure}

Following~\cite{piet18} we next show that an approximation of the normative model given by Eq.~\eqref{eq:ode_changing} can be tuned to give near optimal accuracy, but the accuracy of the approximation tends to be sensitive to the changes in the discounting parameter.  This approximate, linear model is given by,
\begin{equation}
\frac{dy_t}{dt}=\kappa \left[ \sum_{i=1}^{\infty} \delta(t-t^R_i)- \sum_{j=1}^{\infty} \delta(t-t^L_j)\right]-\gamma\cdot y_t.
\label{eq:ode_linear}
\end{equation} 
In particular, here the nonlinear $\sinh$ term in Eq.~\eqref{eq:ode_changing} is replaced by a linear term proportional to the accumulated evidence.

When tuned appropriately, Eq.~(\ref{eq:ode_linear}) closely approximates the dynamics and accuracy of the optimal model (Fig.~\ref{fig3_linear}A)~\citep{piet18,velizcuba16}. Moreover, it also provides a good fit to the responses of rats on a dynamic clicks task~\citep{piet18}. 
As the normative and linear models exhibit similar dynamics, it appears that they are difficult to distinguish.  However, as we show next, the linear model is more sensitive to changes in its discounting parameter, providing a potential way to distinguish between the models.


We assume that $T$ is large enough so that accuracy has saturated (as in Fig. \ref{fig1_snr}C), and compare the maximal accuracy of the nonlinear and linear model. For the linear discounting rate  that maximizes accuracy, $\gamma = \gamma^*,$  the linear model obtains accuracy nearly equal to the normative model (Fig.~\ref{fig3_linear}A, inset).  The optimal linear discounting rate, $\gamma^*,$ increases with SNR (Fig.~\ref{fig3_linear}A), whereas the discounting term in the normative, nonlinear model remains constant when the hazard rate, $h,$ is fixed. When SNR is large, evidence discounting in the linear model can be stronger (larger $\gamma^*$), since each evidence increment is more reliable and can be given more weight.
When SNR is lower, linear evidence discounting is weaker (smaller $\gamma^*$) resulting in the averaging of noisy evidence across longer timescales.

What is the impact of using the wrong (suboptimal) evidence discounting rate in the two models? To answer this question we  compare the accuracy of two observers, one using the nonlinear model with a wrong hazard rate, $\tilde{h} \neq h$, and the other using the linear model with a suboptimal discounting rate $\tilde{\gamma} \neq \gamma^*$.  As shown in Fig.~\ref{fig3_linear}B accuracy is more sensitive to relative changes in $\tilde{\gamma}$ in the linear model, than relative changes  in the assumed hazard rate, $\tilde{h},$ in the nonlinear model. We quantified the sensitivity of both models to changes in evidence discounting rates by computing the curvature of accuracy functions at the optimal discounting value over a range of SNRs (Fig.~\ref{fig3_linear}C).

Both models are insensitive to changes in their discounting parameter at low SNR (bottom curve of Fig.~\ref{fig3_linear}B). This result is intuitive, as when SNR is small observers perform poorly regardless of their assumptions.  On the other hand, when SNR is high  observers receive strong evidence from a single click, and the nonlinear model adequately adapts across a broad range of discounting parameter values. The linear model, however, is still sensitive to changes in the discounting parameter, $\gamma$. At high SNR, the belief, $y_t,$ as descried by either model is driven to larger values. Whereas the nonlinear model can rapidly discount extreme beliefs as it includes a supralinear leak term, the linear model is not as well adapted, and requires fine tuning. Note, however, that at values of SNR  higher than the ones used in Fig.~\ref{fig3_linear}B,  when, for instance, a single click is sufficient for an accurate decision, both the linear and nonlinear models are insensitive to changes in their discounting parameters. 
 We also note that the insensitivity of the nonlinear model to changes in the discounting rate, $h$, suggests that this is a more robust model: An observer who does not learn the hazard rate, $h$, exactly can still perform well.  A linear model requires finer parameter tuning to achieve maximal accuracy.
 
 The nonlinear model obtains maximal accuracy as long as the assumed hazard rate matches the true hazard rate $\tilde{h} = h$. On the other hand, the optimal discounting rate of the linear model is also sensitive to changes in the SNR due to changes in the click rates. To quantify this effect, we computed the ratio between the maximal accuracy of the linear model with discounting rate $\tilde{\gamma}$ to the maximal accuracy of the nonlinear model with $\tilde{h} = h$ as the SNR was varied, but $h$ was kept fixed (Fig.~\ref{fig3_linear}D). To compute the maximal accuracy we kept $\tilde{\gamma}$ fixed at $\gamma^*$, the optimal discounting rate for a reference SNR. The maximal-accuracy ratio for the linear model decreases as SNR changed from this reference SNR, as the  optimal discounting parameter of the linear model depends on SNR, and the hazard rate $h$. Thus, the linear model can achieve maximal accuracy very close to that of the nonlinear model, but this requires fine tuning.


This points to a general difficulty in distinguishing models subjects could use to make inference: Simpler approximations may predict performance that is near identical to that of a normative model. However, this may require precise tuning of the approximations.  If the parameters of the task are changed to differ from those on which the subjects have been trained, \emph{i.e.} on tasks where subjects are lead to assume incorrect parameters, the normative model may behave differently from the approximations.  In the case we considered, the models may be distinguishable if an animal is extensively trained on trials with fixed parameters $h$ and $\S,$ but subsequently interrogated using occasional trials with different task parameters.

The preceding point is illustrated by the following thought experiment. Assume a subject is extensively trained on a fixed set of task parameters: $\mathcal{S}^\text{ref}=2$, $(\lambda_\text{low}, \lambda_\text{high})=(2, 8.5)$Hz, $h=1$Hz (peak of the red curve in Fig.~\ref{fig3_linear}D). We then introduce some trials with different click rates, say
$\mathcal{S}^\text{new}=5$ with $(\lambda_\text{low}^\text{new}, \lambda_\text{high}^\text{new})=(12,53)$Hz 
and $h=1$Hz, chosen so that $\kappa=\log (\lambda_\text{high}/\lambda_\text{low})$ is constant across the two conditions. 
We denote by Acc$_\text{lin}(\mathcal{S})$ and Acc$_\text{norm}(\mathcal{S})$ the accuracies of an observer using the linear and normative models on trials with a given $\mathcal{S}$. Since the subject was trained on click rates that correspond to $\mathcal{S}^\text{ref}$, their discounting strategy will be adapted to these values. Note that the ratio between Acc$_\text{norm}(\mathcal{S})$ and Acc$_\text{lin}(\mathcal{S})$ when $h = 1$ is the red curve in Fig.~\ref{fig3_linear}D. Since the ratio between Acc$_\text{norm}(\mathcal{S}^\text{ref})$ and Acc$_\text{lin}(\mathcal{S}^\text{ref})$ is near 1, the linear and normative models cannot be distinguished at $\mathcal{S}^\text{ref}$.
However, a subject using the normative model tuned at $\mathcal{S}^\text{ref}$, will still perform optimally at  $\mathcal{S} \neq  \mathcal{S}^\text{ref}$, if $\kappa$ and $h$ are held constant.
On the other hand, a linear model optimized at $\mathcal{S}^\text{ref}$, will no longer be optimal at $\mathcal{S} \neq  \mathcal{S}^\text{ref}$. This distinction is captured by the drop in the accuracy ratio along the red curve in Fig.~\ref{fig3_linear}D. \\
\indent
We can quantify the distinction between the two models by their relative difference:
\begin{equation*}
\frac{\text{Acc}_\text{norm}(\mathcal{S}^\text{new})-\text{Acc}_\text{lin}(\mathcal{S}^\text{new})}{\text{Acc}_\text{norm}(\mathcal{S}^\text{new})}=0.05.
\end{equation*}
More generally, for any decision making model, we may define the quantity
\begin{equation*}
D_\text{model}(\mathcal{S}):=
\frac{\text{Acc}_\text{norm}(\mathcal{S})-\text{Acc}_\text{model}(\mathcal{S})}{\text{Acc}_\text{norm}(\mathcal{S})},
\end{equation*}
which will equal 0 if the model used is the normative one.  
If we compute $D_\text{model}(\mathcal{S})$ using responses from a real subject, one can generate curves such as those in Fig.~\ref{fig3_linear}D. If the curves are not constant (equal to 1), this would suggest the subject is not using an optimal model. Furthermore, a single value of $\mathcal{S}^\text{new}$ for which $D_\text{model}(\mathcal{S}^\text{new})\neq 0$ provides evidence that the model is not optimal.
In the next two sections, we show how the linear and nonlinear model with added sensory noise differ when  fitting the discounting parameters to choice data.


\section{Fitting discounting parameters in the presence of sensory noise}
\label{sec:identify}

The models we have discussed so far translate sensory evidence  into decisions deterministically, 
and do not account for the nervous system's inherent stochasticity~\citep{Faisal2008}.  We next asked whether the inclusion of sensory noise leads to further differences between the two models, particularly when fit to choice data.

\cite{Brunton2013} showed that in the static version of the clicks task humans and rats make decisions that are best 
described by a model in which evidence obtained from each click
is variable. In the dynamic 
 version of the task, \cite{piet18} showed that rats' suboptimal accuracy is well explained by a model that includes similar internal variability. 
\cite{piet18}  modeled such ``sensory'' noise either by applying Gaussian perturbations to
the evidence pulses, or by attributing, with some mislocalization probability, a click coming from the right or left speaker to the wrong side. 

As a minimal model of neural or sensory noise, we too introduced additive Gaussian noise into the evidence pulse of each click, so that the nonlinear model in Eq.~\eqref{eq:ode_changing} takes the
form
\begin{equation}
\D\frac{dy_{t}}{dt}=\sum_{i=1}^\infty
\eta_i\delta\left( t-\rcli{i}\right)-
\sum_{j=1}^\infty
\zeta_j\delta\left( t-\lcli{j}\right)
-2h\sinh(y_{t}),
\label{eq:sinh_clicks_noisy}
\end{equation}
where $\eta_i,\zeta_j \sim {\mc N}(\kappa, \sigma)$ are i.i.d.\ Gaussian 
random variables with mean $\kappa$
and standard deviation $\sigma$. 
Similarly,
the linear model from Eq.~\eqref{eq:ode_linear} becomes:
\begin{equation}
\frac{dy_{t}}{dt}=\sum_{i=1}^\infty\eta_i
\delta\left(t-\rcli{i}\right)-\sum_{j=1}^\infty\zeta_j
\delta\left(t-\lcli{j}\right)-\gamma  
y. \label{eq:noisy_linear}
\end{equation}

\begin{figure}[!b]
	\includegraphics[width=\textwidth]{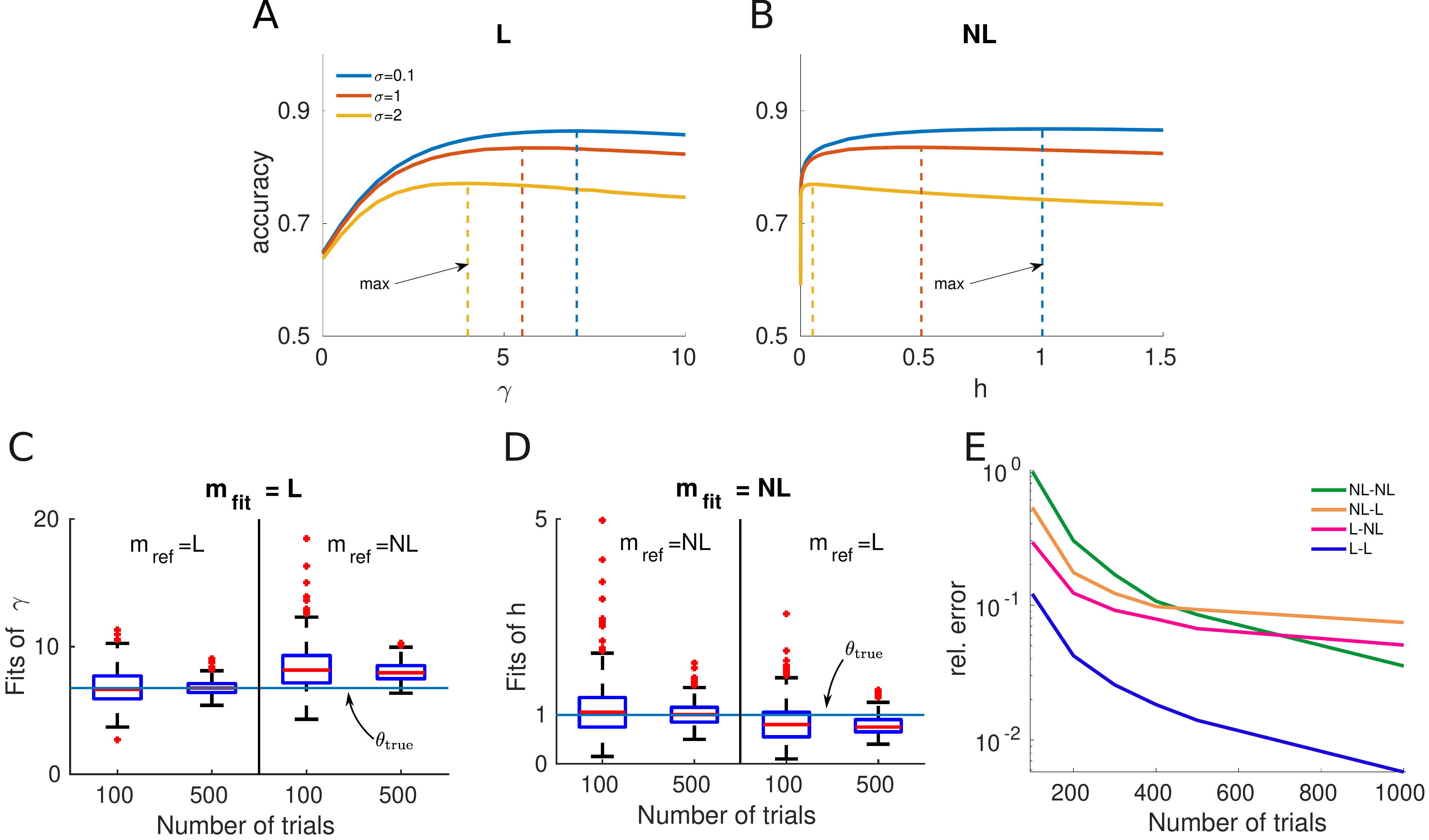}
	\caption{{\bf A:} Accuracy as a function of the discounting parameter $\gamma$ for the linear model with sensory noise  described by Eq.~(\ref{eq:noisy_linear}).  As noise increases, maximal accuracy is achieved at lower
	discounting values (dotted lines).  {\bf B:} Same as A for the nonlinear model with discounting parameter $h$.
	{\bf C,D}
	The whisker plots represent the spread of the posterior modes (MAP estimates) obtained across the 500 fitting procedures, for each model pair and reference dataset size. 
	On each box, the red line indicates the median estimate.  The MAP estimates are closer with more trials, but are biased in the case of model mismatch.  
	{\bf E:} Average relative error, Eq.~(\ref{eq:error}), in fitting the discounting parameter
	as a function of reference dataset size.
	Each color corresponds to a specific pair
	$(m_\text{fit},m_\text{ref})$. In the case of model mismatch (L-NL and NL-L curves)  the relative error will not converge to zero due to the bias in the parameter estimate. Panels C-E describe the same set of fits. See Appendix~\ref{app:mc} for simulation details.
}
	\label{fig4}
\end{figure}

Before  fitting these models to choice data, we note that
an increase in sensory noise, $\sigma$, decreases the value of the discounting parameters that maximize accuracy
in both models~\citep{piet18}: Noisier observations require integration of information over longer timescales  (Fig.~\ref{fig4}A,B). 
Thus, adaptivity to change points is sacrificed in order to pool over larger sets of observations .  This, in turn, leads to
larger biases, particularly after change points. 
A similar trade-off between adaptivity and bias 
has been observed in models and human subjects performing a related dynamic decision task~\citep{glaze18}.

We next fit the discounting parameters in both models using synthetic choice data, treating the other parameters of the models as known. To do so we 
produced responses using a fixed  \emph{reference model} from both classes, and fit a model  from each class to the resulting datasets. 
Specifically, let $m_\text{ref}\in\{\text{L},\text{NL}\}$ (L = linear, NL = nonlinear) denote the reference model 
used to produce the choice data, and let $m_\text{fit}\in\{\text{L},\text{NL}\}$ denote the 
model that was fit to the resulting data. We independently studied the four possible model pairs $(m_\text{fit},m_\text{ref})$. In what follows, $\theta$ refers to the discounting parameter that was fit
to data in any given class, so that $\theta:=\disclin$ when $m_\text{fit}=\text{L}$ and $\theta:=\hzd$ when $m_\text{fit}=\text{NL}$. 
Note also that the hazard rate parameter, $\theta = h,$ that was fit to data in the case $m_\text{fit}=\text{NL}$ is distinguished both 
from the hazard rate, $h_\text{stim},$ used to generate click stimuli, and the hazard rate, $h_\text{ref}$, used
to produce the reference choice data of the nonlinear model.  
Therefore, to remove ambiguity, we denote by $h_\text{ref}, \gamma_\text{ref}$ the two constant discounting
parameter values used to produce the reference choice data with the nonlinear and linear
models, respectively.  
To pick these constants in our simulations, we  took the 
values that would maximize accuracy in the corresponding noise-free systems. That is, 
$h_\text{ref} := h_{\rm stim} =  1$ and $\gamma_\text{ref} := \gamma^* \approx 6.75$ (See 
Appendix~\ref{app:mc} for more details on the simulations).

During a single fit, we generated
stimulus data for $N$ i.i.d.\ trials,
\begin{equation}
\Data:=\LE\{\LE(\train_k, d_k\RI):1\leq k\leq N\RI\},
\label{eq:data}
\end{equation}
where $\train_k:=\left( \left\{ t_i^R \right\}_{i=1}^{N_{R,k}},\left\{ t_j^L \right\}_{j=1}^{N_{L,k}} \right)$ is the sequence of $N_{R,k}$ right clicks and $N_{L,k}$ left clicks on trial $k$, and
$d_k\in\{0,1\}$ is the choice datum for this trial.
We used Bayesian parameter estimation (See Appendix~\ref{appendix:fitting} for details) to obtain a posterior probability distribution over the discounting parameter, $\PP(\theta|\Data)$. 

To account for the variability in the posteriors that arise due to finite size effects, we performed 
$M=500$ independent fits  per model pair $(m_\text{fit},m_\text{ref})$, with different dataset sizes: 
$N\in\{100,200,300,400,500\}$.
To quantify the goodness of these fits, we used the relative mean posterior squared error,
averaged across the $M$ fits,
\begin{equation}
\text{err}\LE(p_1,\ldots,p_M,\theta_\text{true}\RI)
:=\frac{1}{\theta_\text{true}^2 M}
\sum_{i=1}^M\int_0^\infty 
p_i(\theta) (\theta-\theta_\text{true})^2
d\theta.
\label{eq:error}
\end{equation}
This quantity provides a relative measure of how close the posterior distribution is to $\theta_{\text{true}}$. Here $p_i(\theta)$ denotes the posterior density, $\PC{\theta}{\Data}$, from fit $i$.
Note, the definition of $\theta_\text{true}$ is nuanced. If the reference and fit models are the same, then $\theta_{\text{true}}$ is set to the ground truth, {\emph i.e.} the discounting parameter value used to produce the reference choice data (e.g., $\theta_{\text{true}} = h_\text{ref}$ when $m_{\rm fit} = m_\text{ref} = \text{NL}$). However, when the fit and reference model classes differ (i.e. when $m_\text{fit}\neq m_\text{ref}$), then there is no obvious ground truth, and $\theta_{\text{true}}$ must be defined differently.
In this case, we used the correspondence $\gamma_\text{ref} \leftrightarrow h_\text{ref}$.   
That is, when fitting the nonlinear model we always set $\theta_{\rm true} := h_\text{ref}$, 
and when fitting the linear model, we always set $\theta_{\rm true} := \gamma_\text{ref}$.
There are other possible ways of defining  $\theta_{\rm true}$ in this case, such as picking a discounting parameter value for the fit model class that produces the same accuracy as the reference model. 
Although arbitrary, our definition is sufficient to illustrate -- as we show next -- that cross-model fits are feasible and that the
$m_\text{fit}=\text{L}$ case is qualitatively different than the $m_\text{fit}=\text{NL}$ case, regardless of the reference model class.
However, due to the model mismatch, we expect a bias in the parameter
estimate for these situations (i.e. an error that does not converge to $0$), unless we define the ground truth self-referentially as the value of the parameter
for which the estimate is unbiased. 

The maximizer $\theta^*$ of our Bayesian posterior $\PP(\theta|\Data)$ defines the maximum likelihood estimate (MLE)\footnote{Which is equal to the Maximum A Posteriori (MAP) estimate in our case, as we picked a uniform prior over a wide interval (See Appendix~\ref{appendix:fitting} for details).} of our discounting parameter. We plot the distribution of these across the 500 independent fits, for each $(m_{\rm fit}, m_{\rm ref})$ pair in Fig.~\ref{fig4}C,D.
As the number of trials used in the reference dataset increased from 100 to 500, the spread of the estimates diminishes. However, a bias in the estimate appears whenever $m_\text{fit}\neq m_\text{ref}$. For reference datasets of size 500, 98\% of the 500 MAP estimates in the L-NL fits 
	lie strictly above $\gamma_\text{true}$, versus 50.4\% for the corresponding L-L fits. Similarly, 86.6\% of the estimates in the NL-L fits 
	lie strictly below $h_\text{true}$, versus 44.2\% for the corresponding NL-NL fits.

We found that the relative error from Eq.~(\ref{eq:error}) decreases as larger blocks of trials are used to fit the discounting parameter (Fig.~\ref{fig4}E). 
We note the following parallels between the sensitivity to parameter perturbation of each model class (explored in Fig.~\ref{fig3_linear}B,C) and the decreasing rate of the relative errors for each model pair.
As expected, a model that produces responses that are less sensitive to changes in its discounting parameter requires more trials to be fit to data: The reduction in relative error is the slowest for the $(\text{NL},\text{NL})$ and $(\text{NL},\text{L})$ pairs. This is consistent with the insensitivity of the nonlinear model to changes in discounting parameter, making it difficult to identify its parameters.
On the other hand, the linear model fits  -- $(\text{L},\text{NL})$ and $(\text{L},\text{L})$ -- converge more rapidly, likely because the linear model is sensitive to changes in its discounting parameter (See Fig.~\ref{fig3_linear}B,C). 

In anticipation of our next section, we  point out that computing the MLE can be treated as a statistical learning problem in which we minimize a negative log-likelihood loss function over the dataset $\Data$~(See Eq. 7.8 in \cite{friedman01}):
\begin{equation}
\mathcal{L}_{LL}(d_k, m_\text{fit}({\mc T}_k) | \sigma, \theta):=-\log \PP (d_k = m_\text{fit}({\mc T}_k)| \sigma, \theta).
\label{eq:llh_loss}
\end{equation}
Here $d_k$ and $m_\text{fit} ({\mc T}_k)$ are the choices generated by the reference and fit models, respectively, on the $k^\text{th}$ trial. As before the discounting parameter, $\theta$, and the level of sensory noise, $\sigma,$ parametrize the fitted model.
Fitted model responses $m_\text{fit} ({\mc T}_k)$ are non-deterministic only because of sensory noise. 
The likelihood $\PP (d_k = m_\text{fit}({\mc T}_k) | \sigma, \theta)$ is the probability that the response generated by the fit model on trial $k$ matches the response observed in the data (See Appendix~\ref{app:mc} for details on how this likelihood was computed for each model class), which must be obtained from many realizations of $m_\text{fit} ({\mc T}_k)$ subject to click noise of amplitude $\sigma$.
The MLE, $\theta^*,$ for $m_\text{fit}$ is then found by minimizing the expected loss 
across all trials, 
\begin{align}
\theta^* :=  {\rm argmin}_{\theta} {\rm E} \left[ \mathcal{L}_{LL}(d_k, m_\text{fit}({\mc T}_k) | \sigma, \theta) \right] =  {\rm argmin}_{\theta} \frac{1}{N} \sum_{k=1}^N \mathcal{L}_{LL}(d_k, m_\text{fit}({\mc T}_k) | \sigma, \theta), \label{lossminprob}
\end{align}
taking the expectation over all $N$ samples in the dataset, but conditioning on the fitted model's discounting parameter, $\theta$ and noise amplitude, $\sigma$. As the MLE is consistent, we expect the fit parameters will converge to the true parameters~\citep{Wald1949} (Fig.~\ref{fig4}C,D). Framing Bayesian parameter estimation  in this way will help us compare to our approach of fitting by minimizing the 0/1-loss function we introduce next.




\section{Fitting with the 0/1-loss function}
\label{S:fitting}


We next asked how the parameters that define the model whose responses best match the choices of a reference observer compare to those that maximize the likelihood of observing these choices.
As we noted minimizing the log-likelihood loss ${\mc L}_{LL}$ given in Eqs.~(\ref{eq:llh_loss}) and (\ref{lossminprob}) gives the parameters most likely to have produced the data, and we expect the corresponding estimates of the discounting parameter to converge to the true value when the fit and reference models match.

To find the parameters that maximize the probability of matching the choices of the model to those observed in a dataset on every trial, we define the 0/1-loss function, 
\begin{equation*}
\mathcal{L}_{0/1}(d_k,m_{\text{fit}}({\mc T}_k, Z_j) | \sigma, \theta )
:=\mathbb{1}(d_k \neq m_{\text{fit}}(\mathcal{T}_k, Z_j) | \sigma, \theta),
\end{equation*}
where $\mathbb{1}$ is the indicator function, $({\mc T}_k, d_k)$ is a data sample indicating the click stimulus and response on a trial $k$, and $m_{\text{fit}}(\mathcal{T}_k, Z_j)$ is the response of the fitted model with discounting parameter $\theta$, click stimulus ${\mc T}_k$, and  $Z_j$,  $j=1,...,Q$ are realizations of sensory noise, \emph{i.e.} a sequence of i.i.d. Gaussian variables that perturb the evidence obtained from each click. We will marginalize over realizations of the (unobserved) sensory noise, and $Q$ denotes the number of realizations we use in the actual computation. Fitting the discounting parameter $\theta$ then involves minimizing the empirical expectation of the loss function ${\mc L}_{0/1}$ over the data samples $({\mc T}_k , d_k)$ and across realizations, $Z_j,$ of sensory noise,
\[
\theta^* : = {\rm argmin}_{\theta} {\rm E} \left[ \mathcal{L}_{0/1}(d_k, m_{\text{fit}}({\mc T}_k, Z_j)  | \sigma, \theta) \right] = {\rm argmin}_{\theta} \frac{1}{QN} \sum_{j=1}^Q \sum_{k=1}^N \mathcal{L}_{0/1}(d_k,m_{\text{fit}}({\mc T}_k, Z_j)  | \sigma, \theta).
\]
For a binary decision model, this involves finding the parameter $\theta$ that minimizes the expected number of mismatches (or probability of a mismatch) between the choices of the model and those observed by the data (minimizing 0/1-loss), or maximizes the expected number of matches (or probability of a match) between the data and fit model (maximizing 0/1-prediction accuracy). In our fits, we used $Q=1$, sampling a single realization of click noise perturbations per click stream. As we sampled from a large number of click streams, this was sufficient to average the loss function.

Both loss functions, $\mathcal{L}_{LL}$ and $\mathcal{L}_{0/1}$, are commonly used  to fit models to data~\citep{friedman97,friedman01}. Minimizing the expectation of $\mathcal{L}_{0/1}$ is reasonable, as it seems likely that the parameters that define the model that matches the largest number of choices observed in the data should be close to the one the reference observer actually uses (assuming that there is no model mismatch).  These parameters will then also  best predict future responses.  On the other hand, minimizing $\mathcal{L}_{LL}$ produces the most likely parameters that produced the observed data.  

However, it is well-known that parameters estimated using different cost functions can differ, even when the models used to fit and generate the data agree.  To see the difference between  using $\mathcal{L}_{LL}$ and $\mathcal{L}_{0/1}$ in Eq.~(\ref{lossminprob}) consider a Bernoulli random variable, $B$ with parameter $p >0.5$.  Given a large sequence of observed outcomes, $N \to \infty$, the parameter that minimizes the expected loss $\bar{\mc L}_{LL}$ converges to $p,$ as the MLE is consistent and asymptotically efficient~\citep{Wald1949}.  On the other hand, the parameter that minimizes the expected loss $\bar{\mc L}_{0/1}$ is $p = 1$ (See Appendix~\ref{app:bernoulli}): The individual outcomes in a series of independent trials are best matched by a model that always guesses the more likely outcome.  

We observed a similar bias when we used the  $\mathcal{L}_{0/1}$ loss function to infer the discounting parameters in our evidence accumulation models~\citep{friedman97}: We generated a set of $10^6$ click-train realizations, ${\mc T}_k,$ and two sets of responses, $d_k,$ from each the linear and nonlinear evidence accumulation models with sensory noise (See Appendix \ref{app:mc}).
Next we used these stimulus realizations  as input to an evidence-accumulation model (linear or nonlinear) with  a fixed discounting parameter to produce a corresponding set  of $10^6$   \emph{reference observer responses}.  We generated a second set of \emph{model responses} using the same database of $10^6$ click-train realizations, but allowed the discounting parameter $\theta$ to vary.  We call the fraction of time the reference observer and model responses agree the \emph{0/1-prediction accuracy} (PA) of the model, the complement of the expected 0/1-loss over a test set, ${\rm PA} = 1 - \bar{\mc L}_{0/1}$.
When the model and reference observer agree the PA is 1 in the absence of sensory noise ($\sigma = 0$), as the stimulus determines the choice fully.  However, PA decreases as sensory noise increases.


\begin{figure}[!t]
{
	\centering
\includegraphics[width=13cm]{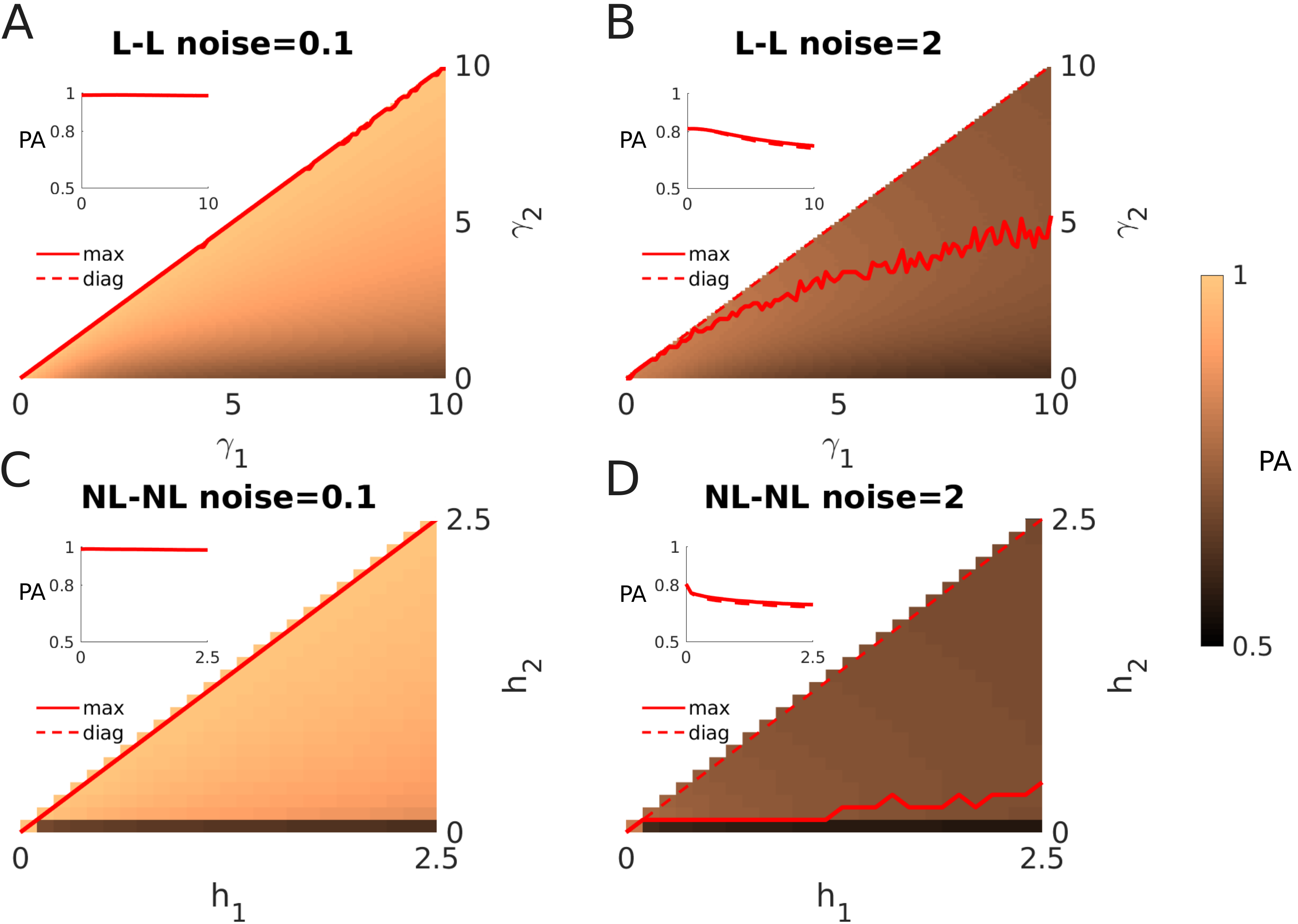}
	\caption{
	The 0/1-prediction accuracy (PA) of a fit model as a function of the discounting parameter of the fit model (vertical axis), and model used to generate training data (horizontal axis). Each column corresponds to a noise value, and each row to a model (linear or nonlinear).  Colors indicate the variation of the PA as the discounting parameters are varied (for $\gamma_2 \leq \gamma_1$, as PA is symmetric about the diagonal). Red curves represent the fit parameter that maximizes percent match, as a function of the reference parameter. For higher noise values, this lies well below the diagonal, $\gamma_1 = \gamma_2$, which would correspond to matching the parameters of the reference model and data. The same $10^6$ click realizations were used across all panels, but each decision from the two models was computed with independent sensory noise realizations. Other parameters are $h_{\text{stim}}=1$ Hz, $(\lambda_{\text{high}},\lambda_{\text{low}})=(20,5)$ Hz, and $T=2$ s.}
\label{fig:PP}
}
\end{figure}


Somewhat surprisingly, the parameters that minimize expected 0/1-loss are biased, and this bias increases with sensory noise (Fig.~\ref{fig:PP}).
In particular, the discounting parameter that best predicts the reference observer responses  is lower than the one used to generate these responses (Fig.~\ref{fig:PP}B,D). This is consistent with our observations in Section~\ref{sec:identify}, as integration over longer periods of time decreases response variability (Fig.~\ref{fig4}A,B).
This tendency is pronounced when larger values of the discounting parameters are used to produce the training data. Larger discounting leads to shorter integration time, and  increased variability in the responses. Furthermore, the nonlinear (NL) model exhibits this bias much more strongly than the linear model (L).
See Appendix~\ref{app:bias} for a possible metric of the reported bias, and its dependence on sensory noise for each model class (Fig.~\ref{fig:bias_metric}).

Thus sensory noise is the main reason the expected 0/1-loss is minimized at a discounting parameter that does not match the one used to generate the data. Such internal noise introduces variability in the responses: even the same model will not match its own responses given the same stimulus, and a decrease in output variability can increase the PA of a model. In the present case, such a decrease in response variability is achieved by decreasing the  discounting parameter, and increasing integration time.

We expect that similar biases occur whenever a 0/1-loss function is used to fit models to choice data. Sensory noise, lapses in attention,
and numerous other sources of noise nearly always introduce some variability in the responses of observers.  In such cases, models that are less variable than the observer may best match an observed set of responses, and best predict future responses~\citep{friedman01}.  
However, these parameters are not always most likely to have been used by the observer.  Using a 0/1-loss function may thus not always reveal the process that the observer used to generate the responses, even if the model the observer uses is close to the one used to fit the data.


\section{Discussion}

Normative models of decision-making make concrete predictions about the computations and actions of experimental subjects, and can be used to interpret behavioral data~\citep{Geisler03}. 
Such models can also be used to identify task parameter ranges in which observers' responses are most sensitive to their assumptions about the task. 
In turn, such information can  then be used to tease apart candidate model classes the experimental subject might be employing. Here we have focused on properties of a normative, nonlinear model, and its differences with a close, linear approximation. We found that the linear model is more sensitive to changes in the discounting parameter compared to the nonlinear model, and suggest this is why fitting a linear model to choice data requires fewer trials than fitting a nonlinear model.

In dynamic environments, task parameters may have predictable effects on subjects' overall accuracy and accuracy relative to change points. We have shown that there is a range of intermediate to high SNR in which the linear model is sensitive to changes in its discounting parameter, but the nonlinear model is not. This suggests this range could be probed to distinguish the evidence accumulation strategies subjects are using. These strategies may also be fit by other approximate models, like accumulators with no-flux boundaries or sliding-window integrators~\citep{wilson13,glaze15,barendregt19}, which can also be sensitive to changes in their discounting parameters.

Psychophysical tasks used to infer subjects' decision-making strategies can require extensive training and data collection~\citep{hawkins15,hawkins15b}. Normative and approximately normative decision-making models diverge most in their response accuracy when tasks are of intermediate difficulty. As we have shown, task difficulty may be controlled by combinations of task parameters representing fewer dimensions than the total number of parameters. Identifying these parameter combinations may be possible by computing the signal-to-noise (SNR) ratio of the stimulus produced by a particular parameter set. However, subjects' responses are also susceptible to noise from sensing and processing evidence, so it is important to extend descriptions of SNR to account for such factors~\citep{Brunton2013}.

Normative models of evidence accumulation and decision-making can be complex, and simpler, approximately optimal strategies may perform nearly as well~\citep{wilson10,glaze18}. 
If such approximate strategies are easier to learn and tune, subjects may prefer them.  \cite{piet18} showed rats' performance on the dynamic clicks task is well fit by a linear discounting model. However, optimal and well-tuned suboptimal strategies may be difficult to distinguish, and this problem is likely to worsen with increasing task complexity and corresponding model dimensionality. We have described possible model-guided task design approaches that may help tease apart similarly performing models. 

The addition of noise in our evidence accumulation models provides an extra parameter that can account for suboptimal performance. What is the best way to distinguishing whether internal noise  or suboptimal evidence accumulation strategies best account for underperformance? One way to do this, as suggested by our model analysis, is to collect sufficient data over trials in which a task parameter was  changed unbeknownst to the observer. 

For purposes of model fitting to experimental data, we expect that trial-to-trial variability can be more faithfully tracked in the dynamic clicks task than in dynamic decision tasks based on the RDMD task. This is due to the relative simplicity of the clicks as evidence sources: They are either on the right or left, although click side and evidence strength can be misattributed~\citep{piet18}. In contrast, dot motion can be estimated in many ways, making it difficult to  interpret which aspects of the stimulus an animal observed, and used as evidence. Spatiotemporal sampling methods may be too spatially coarse and may require fitting filters to each subject, which could change trial-to-trial~\citep{adelson85,park16}. Transforming click times to delta pulses using Eq.~(\ref{eq:ode_changing}) is more straightforward. Thus, the dynamic click task paradigm is a promising avenue for probing evidence accumulation to complement dynamic tasks which are extensions of classic RDMD~\citep{glaze15}. 

The use of discrete evidence tasks does come with caveats. The neural computations underlying visual motion discrimination in non-human primates are well studied~\citep{born05}, and have a significant history of being linked to decision tasks~\citep{gold07}. As a result, there is an extensive literature connecting neural systems for processing visual motion and those involved in decision deliberation~\citep{shadlen96,Roitman2002}. However, only recently have the neural underpinnings of the decisions of rats performing auditory discrimination tasks been examined~\citep{brody16}. Furthermore, mathematical issues may arise in precisely characterizing discounting between clicks, when evidence arrives discretely. Many different functions could lead to the same amount of evidence discounting between clicks, leading to ambiguity in the model selection process.

Parameter identification for evidence accumulation models can be sensitive to the method chosen to fit model responses to choice data~\citep{ratcliff02}. \cite{glaze15} used the approach of minimizing the cross-entropy error function, which measures the dissimilarity between binary choices in the model and the data. \cite{piet18} used a maximum likelihood approach to identify model parameters that most closely matched choice data. This is related to the Bayesian estimation approach we used to fit parameters of the nonlinear and linear models. We obtained similar results by minimizing the expected 0/1-loss, which biases towards less variable models, especially for models with strong sensory noise (Fig.~\ref{fig:PP}).
A more careful approach to fitting model parameters should also consider penalizing more complex models, which would also allow us to distinguish between the nonlinear and linear model.

\cite{glaze18} recently studied the strategies humans use when making binary decisions in dynamic environments whose hazard rates changed across trial blocks. In this case the ideal observer  must infer both the state, and the rate at which the environment is changing~\citep{Radillo2017}. Interestingly, \cite{glaze18} found that the model that best accounted for response data was not the full Bayes optimal model, but rather a sampling model in which a bank of possible hazard rates replaces the full hazard rate distribution.
Such sampling strategies can more easily be implemented in spiking networks~\citep{buesing11}, and may also arise when considering an information bottleneck, which forces a balance between information required from the past and model predictivity~\citep{bialek01}. As in Occam's razor, the brain may favor simpler models, especially when they perform similarly to more complex models~\citep{balasubramanian97}.

Analyses of normative models for decision-making are important both for designing experiments that reveal subjects' decision strategies and for developing heuristic models that may perform near-optimally~\citep{velizcuba16,piet18,glaze18}. Our findings suggest subjects should be tested mainly at intermediate levels of SNR to provide informative response data. We found that such a level of SNR is between 1 and 2 for an optimal observer, and between 3 and 4 for an observer that uses linear discounting. Tasks that are too easy or hard allow subjects to obtain similar performance with a wide variety of strategies. 
Interestingly, we also found that the models that best predict observer responses, are not necessarily those closest to the ones that the observer is using. 
Moreover, modifications of normative models can also suggest more revealing experiments, like those that include feedback or signaled change points. Ultimately, data from decision-making tasks that require subjects to accumulate evidence adaptively will provide a better picture of how organisms integrate stimuli to make choices in the natural world.

\appendix

\section{Normative evidence accumulation for dynamic clicks}
\label{app:changing}
In dynamic environments, the state $x_t\in\left\{x^R,x^L\right\}$ evolves according
to a continuous-time Markov chain with symmetric transition rates given by the hazard rate, $h$.
We construct a sampled-time approximation $\left\{\tilde{x}_t^{\Delta t}\right\}_{t\in[0,T]}$ of 
the continuous-time Markov process $x_t$, parameterized by $\Delta t$, which is valid for 
$\Delta t$ small enough~\citep{gardiner09}. More precisely, we define a discrete-time Markov
chain $x_n\in\{x^R,x^L\}$ by the transition probabilities: 
$\PP(x_n\neq x_{n-1}|x_{n-1})=h\cdot \Delta t$ and $\PP(x_n=x_{n-1}|x_{n-1})=1-h\cdot \Delta t$,
for all $n\in \mathbb{N}$ and initial condition $x_0$. 
Note that these probabilities are a truncation to first order in $\Delta t$ of the transition 
probabilities that one would otherwise obtain for the embedded discrete-time Markov chain
$\{x_{n\Delta t}\}_{n\in\mathbb{N}}$.
Then, we set 
\begin{equation}
\tilde{x}_t^{\Delta t}:= x_n,\label{eq:sampled-time}
\end{equation}
for all $n\in\mathbb{N}$ and all $n\Delta t\leq t<(n+1)\Delta t$.
In the following, our discrete-time evidence accumulation equations are embedded in 
continuous-time via the correspondence given by Eq.~\eqref{eq:sampled-time}.
As $\Delta t \to 0$ the 
resulting equations apply to the original state process $x_t$ in virtue of the 
sampled-time approximation just described.

Just as in Eq.~\eqref{eq:LPOR}, the log-posterior odds ratio in discrete-time is:
\begin{equation*}
y_n:=\log\frac{\PP(x_n=x^R|s^R(n),s^L(n))}{\PP(x_n=x^L|s^R(n), s^L(n))}.
\end{equation*}
Hence, equations (A.3) and (B.1) from the appendix of~\cite{velizcuba16} hold in our context:
\begin{equation*}
y_n-y_{n-1}=\log\frac{f_{\Delta t}^R(\xi_n)}{f_{\Delta t}^L(\xi_n)}+
\log\frac{1-h\cdot\Delta t+h\cdot\Delta t\cdot e^{-y_{n-1}}}{1-h\cdot\Delta t+h\cdot\Delta t\cdot e^{y_{n-1}}}.
\end{equation*}
In addition, we use the approximation $\log(1+z)\approx z$ for small $|z|$, since $0<\Delta t \ll 1$, so that:
\begin{equation*}
\Delta y_n=\log\frac{f_{\Delta t}^+(\xi_n)}{f_{\Delta t}^-(\xi_n)}-2h\Delta t\sinh(y_{n-1}).
\end{equation*}
Taking the limit $\Delta t\to 0$ yields the ODE:
\begin{align*}
\frac{dy_t}{dt}=\kappa \left[ \sum_{i = 1}^{\infty} \delta(t-t^R_i)- \sum_{j=1}^{\infty} \delta(t-t^L_j)\right]-2h\sinh(y_t),
\end{align*}
or the equivalent rescaled version
\begin{equation*}
\frac{dy_t}{dt}=\sum_{i = 1}^{\infty} \delta(t-t^R_i)- \sum_{j=1}^{\infty} \delta(t-t^L_j) -\frac{2h}{\kappa}\sinh(\kappa\cdot y_t),
\end{equation*}
which both appear in \cite{piet18}.


\section{Derivation of dynamic clicks SNR}
\label{SNR_derive}
Our derivation considers the \emph{signal} in the dynamic clicks task to be the difference in the number of clicks during the final epoch prior to interrogation at time $t = T$. The distribution of final epoch times $\tau$ of the telegraph process $x(t)$ is
\begin{align}
p(\tau) = h \e^{-h \tau} + \e^{- h \tau} \delta ( \tau - T), \hspace{5mm} \tau \in [0,T].  \label{ptau}
\end{align}
The first term is the distribution of waiting times between switches. We  truncate the period at the interrogation time, $T$, and the second term accounts for the probability that no switches occur during the entire trial, and the final and only epoch is of length $T$. For a final epoch of a given length $\tau$, we can describe both the conditional expectation and variance of the difference in click counts $\Delta N$ again using the results of \cite{skellam46}:
\begin{align*}
{\rm E} \left[ \Delta N | \tau \right] = ( \ldh - \ldl) \tau, \hspace{5mm} {\rm Var} \left[ \Delta N | \tau \right] = ( \ldh + \ldl ) \tau.
\end{align*}
Therefore to obtain the unconditional expectation and variance for $\Delta N$, we must marginalize using the laws of total expectation and variance with respect to the distribution of epoch times $\tau$ given in Eq.~(\ref{ptau}). This yields
\begin{align}
{\rm E} [ \Delta N] = (\ldh - \ldl) \int_0^T \tau p(\tau) \d \tau = \frac{1 - \e^{- hT}}{h} \cdot ( \ldh - \ldl) \label{totexp}
\end{align}
for the total expectation. Notice that as $T \to \infty$, the expected number of clicks is limited from above by $\lim_{T \to \infty} {\rm E} [ \Delta N] = (\ldh - \ldl)/h$. Using the law of total variance we can thus compute
\begin{align}
{\rm Var} [ \Delta N] &= {\rm Var} \left[ {\rm E}[ \Delta N | \tau] \right] + {\rm E} \left[ {\rm Var}[ \Delta N | \tau ] \right] = ( \ldh - \ldl)^2 \cdot {\rm Var} [ \tau ] + \frac{1 - \e^{- hT}}{h} \cdot (\ldh + \ldl) \nonumber \\
&= \frac{1 - 2 h T \e^{- hT} - \e^{- 2h T}}{h^2} ( \ldh - \ldl)^2 + \frac{1 - \e^{- hT}}{h} (\ldh + \ldl). \label{totvar}
\end{align}
Plugging Eq.~(\ref{totexp}) and (\ref{totvar}) into the expression for $\SNR_h^T = {\rm E}[ \Delta N] / \sqrt{{\rm Var}[\Delta N]}$ yields
\begin{align}
\SNR_h^T = \frac{(1 - \e^{-hT}) ( \ldh - \ldl) }{\D \sqrt{(1 - 2hT \e^{-hT} - \e^{-2hT}) (\ldh - \ldl)^2 + h \cdot (1 - \e^{-hT}) (\ldh + \ldl)}}. \label{snrht}
\end{align}
Recalling our definition from equation~\eqref{skellam}, 
\begin{align}
{\mc S}:= \frac{\lambda_\text{high}-\lambda_\text{low}}{\sqrt{\lambda_\text{high}+\lambda_\text{high}}}, \label{skellam_appendix}
\end{align}
we can rewrite equation~\eqref{snrht}
in the more convenient form
\begin{align}
\SNR_h^T = F(hT, \S/ \sqrt{h}) = \frac{(1 - \e^{-hT}) \S/ \sqrt{h}}{\sqrt{ (1 - 2hT \e^{-hT} - \e^{-2 hT}) \S^2/h + 1 - \e^{-hT} }},  \label{SNRfin}
\end{align}
where we have highlighted the fact that the SNR is a function of the rescaled trial time $hT$ and the Skellam SNR rate $\S$ scaled by the root of the timescale $\sqrt{1/h}$. Indeed, in the limit as $h \to 0$, we find that $\SNR_h^T \to \SNR_0^T = \S \cdot \sqrt{T}$ consistent with Eq.~(\ref{SNRh0}). We also find that in the limit of infinitely long trials $T \to \infty$, Eq.~(\ref{SNRfin}) tends to
\begin{align*}
\lim_{T \to \infty} \SNR_h^T = \frac{\S/ \sqrt{h}}{\sqrt{\S^2/h + 1}},
\end{align*}
so the SNR is solely determined by $\S / \sqrt{h}$.

Note also that to keep Eq.~(\ref{SNRfin}) constant it is sufficient to keep its constituent arguments constant. This is convenient, since we already must keep $hT$ constant to fix the statistics of information accumulated prior to the final epoch, so we predict that performance is fixed by the following two parameters
\begin{align*}
\frac{\S}{\sqrt{h}} = \frac{\ldh - \ldl}{\sqrt{h(\ldh + \ldl)}} = c_1 \hspace{4mm} \text{and} \hspace{4mm} hT = c_2,
\end{align*}
as reported in Eq.~(\ref{fixparams}).


\section{Diffusion approximation}
\label{app:diffapprox}

Here we demonstrate the diffusion approximation of the normative model for the dynamic clicks task, Eq.~(\ref{eq:ode_changing}) in the limit of large Poisson rates $\lambda_{\rm high}$ and $\lambda_{\rm low}$. Diffusion approximations for jump processes have been addressed by \cite{lansky97}, and \cite{richardson10} who studied the impact of shot noise and pulsatile synaptic inputs on integrate-and-fire models. Following this work, we note that the difference of the click streams in Eq.~(\ref{eq:ode_changing}) can be approximated by a drift-diffusion process with matched mean, $\pm \kappa (\lambda_{\rm hi} - \lambda_{\rm low}),$ and variance, $\kappa^2 (\lambda_{\rm hi} + \lambda_{\rm low})$. This results in the following stochastic differential equation (SDE) for the approximation $\tilde{y}_t$:
\begin{align}
\d \tilde{y}_t = \kappa g(t) (\lambda_{\rm high} - \lambda_{\rm low}) \d t + \kappa \sqrt{\lambda_{\rm high} + \lambda_{\rm low}} \d W_t - 2 h \sinh(\tilde{y}_t) \d t, \label{nldiff}
\end{align}
where $g(t) = \sign \left[ \lambda_R(t) - \lambda_L(t) \right]$ and $\d W_t$ is the increment of a Wiener process. Note the resulting nonlinear drift-diffusion model is similar to the normative models presented in \citep{glaze15,velizcuba16}. The SNR of the signal in Eq.~(\ref{nldiff}) can be associated with the mean divided by the standard deviation in an average-length epoch. Fixing this SNR leads to the relations in Eq.~(\ref{fixparams}). Importantly, the signal in Eq.~(\ref{nldiff}) is characterized entirely by its mean and variance, so we expect that the performance of the model can be directly associated with the SNR. Note, however, that Eq.~(\ref{nldiff}) will only be valid for $\lambda_{\rm high}, \lambda_{\rm low} \gg 1$. Otherwise, one must consider the effects of higher order moments of the click streams, and a prediction of performance purely based on the SNR will break down (Fig.~\ref{fig1_snr}D, Inset), since higher order statistics likely shape response accuracy in these cases.

\section{Model identification}
\label{app:stochmodsim}

We fit parameters of the linear and nonlinear models in two stages. First, we generated synthetic response data from a model (linear or nonlinear) by solving the corresponding ODE or SDE. We then solved a second set of models (linear or nonlinear) for a range of discounting parameters ($\gamma$ for the linear model; $h$ for the nonlinear model), and constructed a posterior distribution over the discounting parameter.
For noisy models, we expect the posterior to be  a smooth function that is peaked around the most likely values of discounting parameter for that trial.
We now describe the details of these parameter fitting procedures for each of the cases: linear vs. nonlinear models.

\subsection{Linear model with stochastic response}
\label{sec:L_stoch_fit}
We incorporate noise into the linear Eq.~(\ref{eq:ode_linear}) by considering multiplicative noise on the click increments, as described by Eq.~(\ref{eq:noisy_linear}).
For a fixed realization of the click train, we can solve this equation explicitly for $y_t$ at the end of  trial $k$:
\begin{align}
y_T^k = \sum_{i=1}^{N_R^k} \eta_i \e^{- \gamma ( T^k - t_R^i)} - \sum_{j=1}^{N_L^k} \zeta_j \e^{- \gamma (T^k - t_L^j)},
\end{align}
where $\eta_i, \zeta_j \sim {\mc N}(\kappa, \sigma)$, revealing $y_{T^k}$ is simply the sum of i.i.d. normal random variables scaled by exponential decay.
Conditioning on the clicks $\T_k$, then $y_{T^k}$ is normally distributed $\left[ y_{T^k}| \T_k, \gamma \right]$ with expectation and variance
\begin{align*}
{\mc E}_k & : = {\rm E} \left[ y_{T^k} | \T_k, \gamma \right] =  \kappa \left[ \sum_{i=1}^{N_R^k} \e^{-\gamma ( T^k - t_R^i)} - \sum_{j=1}^{N_L^k} \e^{- \gamma (T^k - t_L^j)} \right], \\
{\mc V}_k &:= {\rm Var} \left[ y_{T^k} | \T_k, \gamma \right] =  \sigma_c^2 \left[ \sum_{i=1}^{N_R^k} \e^{- 2\gamma ( T^k - t_R^i)} + \sum_{j=1}^{N_L^k} \e^{- 2 \gamma (T^k - t_L^j)} \right],
\end{align*}
so $\tilde{r}_k(\gamma) \in \pm 1$ is a Bernoulli random variable. The likelihood function will be a smooth function of $\gamma$, and determined as an integral over the half-line corresponding to the $\pm 1$ decision:
\begin{align*}
\PP ( \tilde{r}_k = \pm 1  | \gamma , \T_k) &=  \pm \int_0^{\pm \infty} p(y_{T^k} | \T_k , \gamma) \d y_{T^k} = \Phi \left( \pm \frac{{\mc E}_k}{\sqrt{{\mc V}_k}} \right),
\end{align*}
where $\Phi (z)$ is the cumulative distribution function of a standard normal random variable. We can thus compute the posterior over the discounting parameter $\gamma$ as a rescaled product of the likelihoods on each trial.

\subsection{Nonlinear model with stochastic response} 
\label{sec:NL_stoch_fit}
When click heights are noise-perturbed, we cannot explicitly solve the extended nonlinear model. However, we can make progress by applying the idea of mapping between clicks. If we draw trains of clicks, $\T_k$, ahead of time, Eq.~(\ref{eq:sinh_clicks_noisy}) defines the nonlinear model with multiplicative noise.
We can iteratively define the probability density $p(y,t)$ by sampling over the click amplitude noise distribution at each click according to
\begin{subequations} \label{nonlinstoch}
\begin{align}
p(y,t_n^+) &= \frac{1}{\sqrt{2 \pi \sigma^2}} \int_{- \infty}^{\infty} \e^{-(y-z - \kappa_n)^2/2 \sigma^2} p(z,t_n^-) \d z, \\
\frac{\pd p(y,t)}{\pd t} &= 2h \cdot \frac{\pd}{\pd y} \left[ \sinh(y) p (y,t) \right], \hspace{5mm} t_n < t< t_{n+1},
\end{align}
\end{subequations}
where click noise is drawn from the normal distribution ${\mc N}(\kappa, \sigma)$, $t_n$ is the time of the $n$-th click, $\kappa_n = \pm \kappa$ according to the side of the $n$-th click, and we have used the convolution theorem for independent random variables. For any trains of clicks, $\T_k$, Eq.~(\ref{nonlinstoch}) can be solved iteratively to obtain the distribution $p(y,T)$. The likelihood will thus be a smooth function of $h$, determined by the integral over the half-line corresponding to the decision ($\pm 1$):
\begin{align}
\PP (\tilde{r}_k = \pm 1 | h, \T_k) = \pm \int_0^{\pm \infty} p(y_{T^k} | \T_k, h) \d y_{T^k}.
\end{align}


\subsection{Bayesian fitting procedure}
\label{appendix:fitting}
Our goal is to  compute 
or estimate the posterior distribution $\PC{\theta}{\Data}$, which by Bayes' rule is 
proportional to the product of the likelihood of the data with the prior over the 
parameter\footnote{Since all the other task and model parameters are assumed known 
and fixed, we may omit them from the equations.}:
\begin{equation}
\PP({\theta}|{\Data})\propto \PP({\Data}|{\theta})\PP_0(\theta).
\label{eq:post_fit}
\end{equation}
Our method focuses on exploiting the likelihood  function $\PP({\Data}|{\theta})$. We have,
\begin{align*}
\PP({\Data}|{\theta}) =
\PP({\train_{1:N}, d_{1:N}}|{\theta}) =\PP({d_{1:N}}|{\train_{1:N},\theta})\PP({\train_{1:N}}|{\theta}) =\PP({d_{1:N}}|{\train_{1:N},\theta})\PP({\train_{1:N}}),
\end{align*}
where the last step comes from the fact that the  clicks trains are independent of the 
discounting parameter $\theta$ used by the decision-making model\footnote{We remind the 
reader that we operate a distinction between the discounting parameter of the decision
maker and the hazard rate used to produce the data.}. From there, we remark  that the
choice data are conditionally independent  on the clicks stimulus and the discounting parameter. 
Thus,
\[
\PP({d_{1:N}}|{\train_{1:N},\theta})=
\prod_{k=1}^N\PP({d_k}|{\train_k,\theta}).
\]
Therefore we can rewrite  Eq.~\eqref{eq:post_fit} as:
\begin{equation}
\PP({\theta}|{\Data})\propto 
\PP_0(\theta)
\prod_{k=1}^N\PP({d_k}|{\train_k,\theta}).
\label{eq:post_fit_2}
\end{equation}
We use uniform priors for $\theta$, over a finite interval 
$[0,a]$.
In this context, the problem of  computing the posterior distribution of  $\theta$
reduces to assessing the likelihoods of the decision data on each trial, 
$\PP({d_k}|{\train_k,\theta})$ ($1\leq k\leq N$), for a range of $\theta$-values spanning the
interval $[0,a]$. 
In practice, we picked $a=40$ when fitting the linear model and $a=10$ when fitting the nonlinear model.
Finally, note that for numerical stability reasons, our algorithms actually sum 
log-likelihood values, as opposed to multiplying probability values. Relegating the
$\theta$-independent prior into a normalization constant $C$, 
Eq.~\eqref{eq:post_fit_2} becomes, in the log-domain:
\begin{equation}
\log\PP(\theta|\Data)=C+
\sum_{k=1}^N\log\PP(d_k|\train_k,\theta), \quad \theta\in[0,a].
\label{eq:log_post_fit}
\end{equation}


\section{Minimizing 0/1-loss in a Bernoulli random variable}
\label{app:bernoulli}


Consider a simple stochastic binary decision-making model in which we ignore the specifics of evidence sources, as in \cite{pesaran92}. We that  in this case the 0/1-loss function also leads to biased estimates. This result has been pointed out in previous work in which parameter fitting results have been compared between Bernoulli random variables fit with the 0/1-loss function as opposed to maximum likelihood estimators~\citep{friedman97,friedman01}. 

Consider a Bernoulli random variable $B_1$ with success probability $p_1$ generating the reference choices, and the fit Bernoulli model $B_2$ with success probability $p_2$.
Minimizing the log-likelihood loss function recovers $p_2^* = p_1$ in the limit of a large number of trials $N \to \infty$: In this limit, given $p_2$, we have that the expected loss measured by the negative log-likelihood is
\begin{align}
\bar{{\mc L}}_{LL}(d | p_2) = -  \left[ p_1 \log p_2 + (1 - p_1) \log (1-p_2) \right], \label{eq:cross-entropy}
\end{align}
which is minimized\footnote{Note Eq.~\eqref{eq:cross-entropy} is the cross-entropy between $B_1$ and $B_2$.} at $p_2^* = p_1$, the mean of $B_1$. Thus, the parameter from the reference model is recovered, as the Bernoulli random variable satisfies the requirements for the MLE to be consistent~\citep{Wald1949}.

On the other hand, if we fit the parameter $p_2$ by minimizing the expected 0/1-loss function, in the limit of $N \to \infty$ trials, the expected loss is
\begin{align*}
\bar{{\mc L}}_{0/1}(d | p_2) = 1-\PP (B_1 = B_2) = p_1 - p_2(2 p_1-1),
\end{align*}
which decreases in $p_2$ for $p_1 > 0.5$, so the minimal expected loss when $p_1 > 0.5$ is achieved with $p_2 = 1$ (for $p_1<0.5$ it is minimized at $p_2 = 0$).

Of course, the synthetic data and the fit evidence accumulation models we consider are generated from the same click streams on each trial, so a realistic comparison should account for such noise correlations in simplified Bernoulli random variable models, as analyzed in \cite{dai13}.
This analysis is more involved, and we save such a study for future work.

\section{Details on Monte Carlo simulations for figures}
\label{app:mc}

Fig.~\ref{fig1_snr}C was generated using $10^5$ simulations of Eq.~\ref{eq:ode_changing}  from $t=0$ to $t=1$s with the parameters shown in the figure. The time for saturation was chosen to be $0.4$s. For each time between 0 and 0.4s the accuracy was computed as the percentage of the $10^5$ simulations for which the choices were correct. Fig ~\ref{fig1_snr}D was generated using $10^5$ simulations of  Eq.~\ref{eq:ode_changing} for each data point in the $(\ldh, \ldl)$ plane. The maximal accuracy reported corresponds to the numerically computed accuracy at $t=1$s.

Fig.~\ref{fig2_change}B-D was generated using $10^5$ simulations of Eq.~\ref{eq:ode_changing}  from $t=0$ to $t=3s$ with the parameters shown in the figure. The reference change point was chosen to be the last change point in the simulation. For each time between the last change point and one unit of time later, the accuracy is the fraction of the correct responses, simulations for which $\sign (y_T) = x(T)$, the sign of the LLR matched the sign of the telegraph process. Since intervals between change points are exponentially distributed, there are many more data points for short times than for long times after change points. Since some simulations did  not last a full unit of time after the last change point, the number of simulations is less than or equal to $10^5$ (decreasing as time increases). Simulations that had no change point were omitted when computing the accuracy. 

Fig.~\ref{fig3_linear}A was generated as follows. For each value of $\S/\sqrt{h}$, $10^6$ simulations of Eq.~\ref{eq:ode_linear} from $t=0$ to $t=1s$ were generated over a range of $\gamma$ values. For each value of $\gamma$ the accuracy was computed at $t=1s$ and $\gamma^*$ was selected as the value that maximized accuracy. This resulted in a specific value of $\gamma^*$ for each $\S/\sqrt{h}$. Fig.~\ref{fig3_linear}B was generated using $10^6$ simulations of Eq.~\ref{eq:ode_changing} (using $\tilde{h}$ instead of $h$) and Eq.~\ref{eq:ode_linear} (using $\tilde{\gamma}$ instead of $\gamma$) from $t=0$ to $t=1$s for a range of values of $\tilde{h}$ and $\tilde{\gamma}$. For each value of $\tilde{h}$ and $\tilde{\gamma}$, the maximal accuracy was estimated as the value of the accuracy at $t=1$s. Fig.~\ref{fig3_linear}C was generated by estimating the second derivative of the curves shown in Fig.~\ref{fig3_linear}B for each value of $\S/\sqrt{h}$. Fig.~\ref{fig3_linear}D was generated as follows using Eq.~\ref{eq:ode_changing} and Eq.~\ref{eq:ode_linear}. For each of the four curves, $\tilde{\gamma}$ was fixed to the value of $\gamma^*$ corresponding to the reference values $\S/\sqrt{h}=i$ for $i=1,2,3,4$ (see Fig.~\ref{fig3_linear}A). For each curve, this value of $\tilde{\gamma}$ was not changed when new $\S/\sqrt{h}$ values were used. Then, for each curve, the maximal accuracy for the linear and nonlinear models were computed using $10^6$ simulations for a range of new $\S/\sqrt{h}$ values. The quotient of the maximal accuracy of the linear model and the maximal accuracy of the nonlinear model is shown in the figure.

Fig.~\ref{fig4}C-D presents the results of five hundred independent fitting procedures, performed on two different dataset sizes. The parameters for the reference dataset of trials are:
$h_{\text{stim}}=1$ Hz, $(\lambda_{\text{high}},\lambda_{\text{low}})=(20,5)$ Hz, and $T=2$ s.
For each fitting procedure, the trials (either 100 or 500) were sampled uniformly without replacement from a bank of 10,000 trials. The fitting algorithm is an implementation of the Bayesian approach leading to equation~\eqref{eq:log_post_fit} above. When fitting the linear model, the analytical solution from appendix~\ref{sec:L_stoch_fit} was used to compute the likelihood of a single trial ($\PP(d_k|\mathcal{T}_k,\theta)$ term in Eq.~\eqref{eq:log_post_fit}). When fitting the nonlinear model, Monte Carlo sampling was used instead.
More specifically, the distribution of the decision variable at decision time for a given clicks stimulus, $p(y_T | \mathcal{T})$, was estimated by simulating 800 independent trajectories. Thus, each trajectory had its own independent realization of sensory noise but the realization of the stimulus (timing of the clicks) was frozen. Once the density of $y_T$ was estimated, the likelihood term, $\PP(d_k|\mathcal{T}_k,\theta)$ in Eq.~\eqref{eq:log_post_fit}, could be estimated. More details on this method, such as how the number of 800 particles was chosen and how this method was validated on the linear model for which the analytical solution is available, may be found in section 3.5.5 of~\cite{radillo17diss}.

In Fig~\ref{fig4}E, up to trial number 500 on the x-axis, the same fits as in panels C-D were used to compute the relative error (y-axis). Because of the high computational cost of our fitting algorithm (Monte Carlo sampling described above), the points for 1000 trials on the x-axis were computed with only 84 independent fits per model pair (as opposed to 500 for the other points of the figure).

All panels in Fig.~\ref{fig:PP} were produced with a common dataset of $10^6$ trials, generated by presenting the same sets of click streams to the evidence accumulation models.
All trials had same task parameters: trial duration $T=2$s; hazard rate $h=1$ Hz; $\ldh=20$ Hz and $\ldl = 5$ Hz so $\mathcal{S}/\sqrt{h} = 3$; and the initial state of the environment was randomly assigned with a uniform prior. For each panel of Fig.~\ref{fig:PP}, we selected a pair $(m_\text{fit},m_\text{ref})\in \{(\text{L},\text{L}),(\text{NL},\text{NL})\}$ along with a sensory noise amplitude ($\sigma \in \{0.1,2\}$ for $\eta_i, \zeta_i \sim {\mc N}(\kappa, \sigma)$) to be applied to the evidence pulses from the clicks. For each possible pair of discounting parameters ($(\gamma_1,\gamma_2)$ for linear models, $(h_1,h_2)$ for nonlinear models), we computed the $10^6$ decisions (Left or Right) and determined whether the models agreed or not. 
For the linear model, we used values of $\gamma_1, \gamma_2$ between 0 and 10, with increments of 0.1. For the nonlinear model, we used values of $h_1, h_2$ between 0 and 2.5, with increments of 0.1.
For each decision comparison between reference model and fit model, the same click streams were used, but independent noise realizations of click perturbations were applied. The number of agreements was divided by the total number of decision comparisons to produce the color of a single point in the plot.


\section{Bias metric as a function of sensory noise}
\label{app:bias}
\begin{figure}[!h]
	{
		\centering
		\includegraphics[width=13cm]{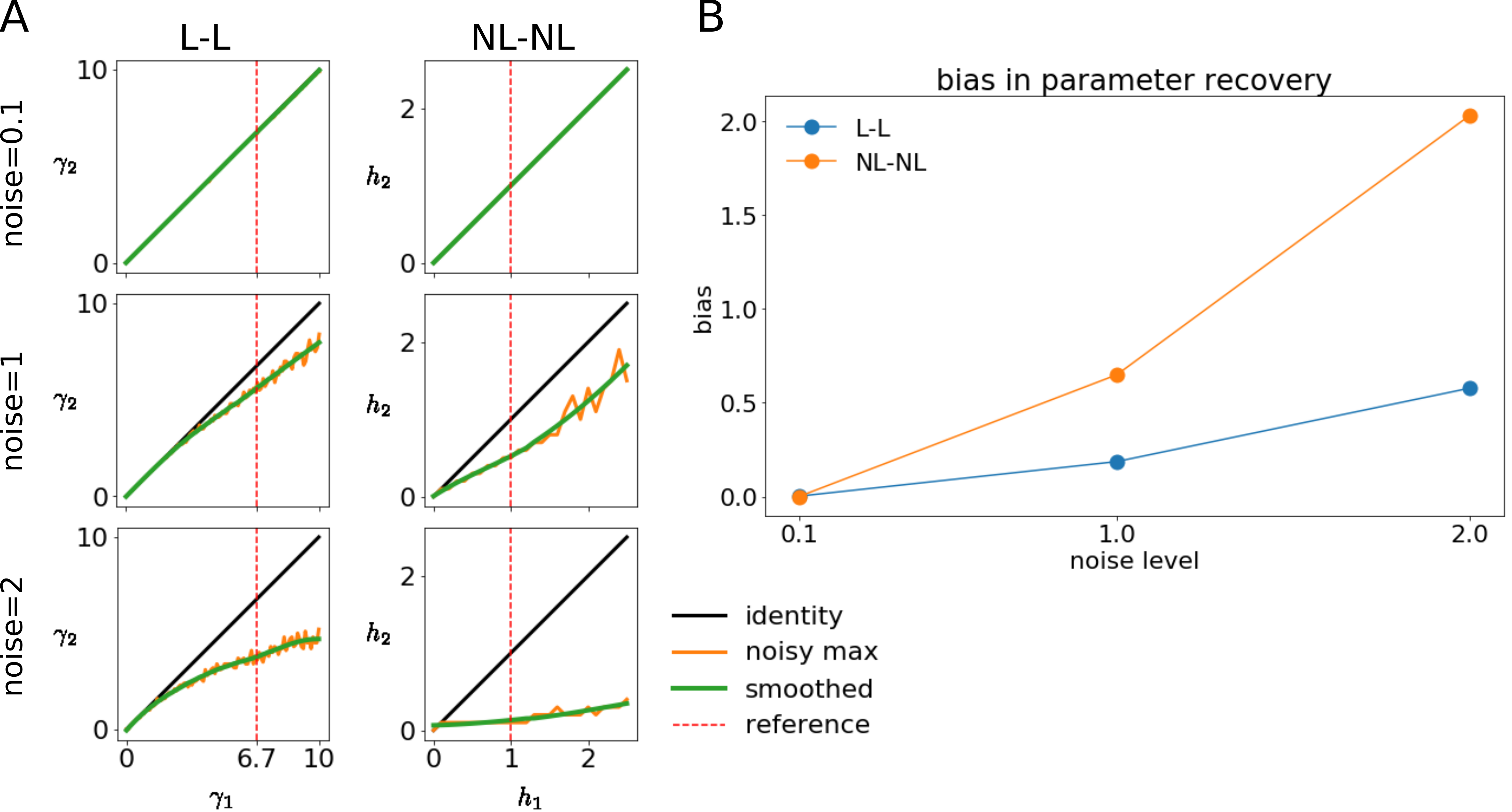}
		\caption{Bias in parameter recovery as a function of sensory noise.
			{\bf A}: Recovered discounting parameter from the fits as a function of the reference discounting parameter used to produce the initial decision data. Top and bottom rows are reproductions of Fig.~\ref{fig:PP} while the middle row is for an intermediate level of sensory noise. The actual fit parameters (golden) were smoothed (green) in order to compute the bias in panel {\bf B} for the reference discounting parameters indicated by the red dotted lines. The black diagonal indicates the identity line, which would correspond to perfect parameter recovery. {\bf B:} Bias in parameter recovery Eq.~\eqref{eq:bias_metric} as a function of sensory noise, for the two model pairs (L-L in blue and NL-NL in golden).}
		\label{fig:bias_metric}
	}
\end{figure}

In this section, we provide additional information about 
the bias in parameter recovery with the 0/1-loss function 
described in Section~\ref{S:fitting}. Fig.~\ref{fig:bias_metric}A includes  results from simulations for noise$=1$ in addition to noise$=0.1$ and noise$=2$ also shown in Fig.~\ref{fig:PP}. 
Bias magnitude and its dependence on sensory noise were determined as follows.
Let $\theta_\text{ref}\in\{\gamma_1, h_1\}$ denote the discounting parameter of the reference model -- this is the model used to produce the initial decision data. Let $\theta_\text{fit}$ denote the
fit value of the discounting parameter, using 0/1-loss minimization. In Fig.~\ref{fig:bias_metric}A, $\theta_\text{ref}$ spans the $x$-axis and $\theta_\text{fit}$
as a function of $\theta_\text{ref}$ is depicted by the golden curve. After smoothing $\theta_\text{fit}$ with a
Savitzky-Golay filter, we obtain $\theta_\text{smooth}$ represented by the green curves in the figure. Picking a fixed reference value for $\theta_\text{ref}$ (red dotted line), we then plot the bias as a function of sensory noise levels in Fig.~\ref{fig:bias_metric}B, where bias is defined as:
\begin{equation}
\text{bias}:=\log\frac{\theta_\text{ref}}{\theta_\text{smooth}}.\label{eq:bias_metric}
\end{equation}
The fixed values of $\theta_\text{ref}$ chosen were the same as in Section~\ref{sec:identify}, $\gamma_1:=6.7457, h_1:=1$.
As described in Section~\ref{S:fitting}, the bias in parameter recovery with the 0/1-loss fitting procedure is more pronounced for the nonlinear model than for the linear model, and increases with sensory noise.

\section*{Acknowledgements}
We thank Gaia Tavoni, Alex Filipowicz, and Alex Piet for helpful feedback on an earlier version of this manuscript. Some computations for this manuscript were done using the  \cite{OhioSupercomputerCenter1987}.




\vfill
\pagebreak
\section*{Reviews and Responses}
\thispagestyle{plain}
\setcounter{page}{1}
\noindent
{\bf Publication Decision 1 from Neurons, Behavior, Data analysis, and Theory} on July 27, 2019 \\[1ex]
\noindent
{\bf Editorial board's determination:} Revise and resubmit \\[1ex]
\noindent
{\bf Comments from the editor.} Sorry again for the long delay. The reviewer is more or less happy with the manuscript, there is a list of suggestions, which I would like to ask you to pay careful attention to before the paper can be accepted. \\[1ex]
\noindent
{\em Reviewers' comments are italicized.} Our responses are in plain text. \textcolor{blue}{Changes to the manuscript are in blue.}  \\
{\bf Comments to the author.} \\
{\em Summary.  This is a careful and thorough paper that presents several results regarding a normative model of decision-making within dynamic environments. The specific case studied is that of an evidence accumulation task presented in Piet et al., 2018. In this task, the subject hears two streams of Poisson-timed clicks coming from their left and right sides. The click rates on both sides may switch during the trial with some hazard rate $h$. The subject is asked to infer which side had the higher rate of clicks when the trial ends, thus making newer information more relevant than older information. In such a task, there are four parameters: $h$, $T$, $\lambda_{\rm high}$, $\lambda_{\rm low}$, where $h$ is the hazard rate, $T$ is the duration of the trial, $\lambda_{\rm high}$ is the higher Poisson rate of clicks, and $\lambda_{\rm low}$ is the lower rate of clicks.

The authors first show that, over a broad range as long as high, low are sufficiently large, the maximal accuracy achievable by the normative model almost exactly depends on only two parameters, ${\mc S}/\sqrt{h}$, and $hT$, where ${\mc S}=(\lambda_{\rm high}-\lambda_{\rm low})/\sqrt{\lambda_{\rm high}+\lambda_{\rm low}}$. They use these two effective parameters to study the model. The authors first compare the behavior of the normative model to a linear model that approximates the normative model, and find that the linear model performs near-optimally if the discounting parameter is finely tuned. To further compare the models, they generated choice data from both the normative model and the linear model, and fit the choices and clicks separately to either the normative model or the linear model. They found that the parameters recovered are biased when the model used to generate data and the model used to fit the data are different. The linear model required less number of trials when it was fit to data; the discounting parameter converged to the true value used to generate data faster for the linear model than for the normative model. The authors note that different cost functions (MLE and 0/1-loss) lead to different estimates of the parameter values and one should be cautious about what cost function to use.

The authors suggest that the normative model and the linear model fits on subjects? performance at different $hT$ and ${\mc S}$ provide a convenient way of identifying the decision-making strategy that the subject is employing in the dynamic clicks task.}\\

\noindent
{\em Major comment:The authors claim that `if an animal is extensively trained on trials with fixed parameters h and S, but subsequently interrogated using occasional trials with different task parameters,' one may be able to identify whether the subject is using the normative model or a suboptimal (or linear approximation) model. This is an important point, and it would be greatly strengthened if this claim can be supported directly with simulations and further elucidated with explanations of the specific steps that should be taken in order to identify the strategies. \\[1ex]
Specifically, in the case where the clicks are generated with $h_{\rm stim}$, and the animal performs normatively with $h_{\rm ref}$, we may fit the normative model and recover $h_{\rm ref}$ reliably. We can compute the maximal accuracy of the normative model given the experimental ${\mc S}/\sqrt{h_{\rm stim}}$, and compare it to the accuracy of the animal. However, if we fit the linear model, how should we go from here? How would one, without knowing $m_{\rm ref}$, and knowing only the clicks/choice data and the experimental parameters $h_{\rm stim}$ and ${\mc S}$, identify the model that the animal is using?} \\[1.5ex]
Indeed, you raise important points, which we can address by taking a closer look at the accuracy ratios we have plotted in Fig.~3D. We now explain the setup of an experiment which could be run to test the sensitivity of a subject's evidence accumulation model to determine whether it is normative-like or more sensitive like the linear models we considered. Of course, in this idealized case, we are not considering internal noise, but the principles would extend to noisy models as discussed in Piet et al (2018). In general, the linear model will not adapt well to interspersed trials with different parameters, whereas the normative model will be more robust. This is now explained in the following paragraphs we have added to the end of Section 4: \\[0.5ex]
\textcolor{blue}{The preceding point is illustrated by the following thought experiment. Assume a subject is extensively trained on a fixed set of task parameters: $\mathcal{S}^\text{ref}=2$, $(\lambda_\text{low}, \lambda_\text{high})=(2, 8.5)$Hz, $h=1$Hz (peak of the red curve in Fig.~3D). We then introduce some trials with different click rates, say
$\mathcal{S}^\text{new}=5$ with $(\lambda_\text{low}^\text{new}, \lambda_\text{high}^\text{new})=(12,53)$Hz 
and $h=1$Hz, chosen so that $\kappa=\log (\lambda_\text{high}/\lambda_\text{low})$ is constant across the two conditions. 
We denote by Acc$_\text{lin}(\mathcal{S})$ and Acc$_\text{norm}(\mathcal{S})$ the accuracies of an observer using the linear and normative models on trials with a given $\mathcal{S}$. Since the subject was trained on click rates that correspond to $\mathcal{S}^\text{ref}$, their discounting strategy will be adapted to these values. Note that the ratio between Acc$_\text{norm}(\mathcal{S})$ and Acc$_\text{lin}(\mathcal{S})$ when $h = 1$ is the red curve in Fig.~3D. Since the ratio between Acc$_\text{norm}(\mathcal{S}^\text{ref})$ and Acc$_\text{lin}(\mathcal{S}^\text{ref})$ is near 1, the linear and normative models cannot be distinguished at $\mathcal{S}^\text{ref}$.
However, a subject using the normative model tuned at $\mathcal{S}^\text{ref}$, will still perform optimally at  $\mathcal{S} \neq  \mathcal{S}^\text{ref}$, if $\kappa$ and $h$ are held constant.
On the other hand, a linear model optimized at $\mathcal{S}^\text{ref}$, will no longer be optimal at $\mathcal{S} \neq  \mathcal{S}^\text{ref}$. This distinction is captured by the drop in the accuracy ratio along the red curve in Fig.~3D. \\
\indent
We can quantify the distinction between the two models by their relative difference:
\begin{equation*}
\frac{\text{Acc}_\text{norm}(\mathcal{S}^\text{new})-\text{Acc}_\text{lin}(\mathcal{S}^\text{new})}{\text{Acc}_\text{norm}(\mathcal{S}^\text{new})}=0.05.
\end{equation*}
More generally, for any decision making model, we may define the quantity
\begin{equation*}
D_\text{model}(\mathcal{S}):=
\frac{\text{Acc}_\text{norm}(\mathcal{S})-\text{Acc}_\text{model}(\mathcal{S})}{\text{Acc}_\text{norm}(\mathcal{S})}
\end{equation*}
which will equal 0 if the model used is the normative one.  
If we compute $D_\text{model}(\mathcal{S})$ using responses from a real subject, one can generate curves such as those in Fig.~3D. If the curves are not constant (equal to 1), this would suggest the subject is not using an optimal model. Furthermore, a single value of $\mathcal{S}^\text{new}$ for which $D_\text{model}(\mathcal{S}^\text{new})\neq 0$ provides evidence that the model is not optimal.
}

\thispagestyle{plain}

\noindent
{\em Minor comments: \\[0.5ex]
Introduction: when the term `nonlinear model' is first used, the authors haven't yet clarified that here the term is synonymous with `normative model.' Also, some readers might not be familiar with what a 0/1 loss function is, making that part of the Intro a little unclear for them.}\\[1.5ex]
It is important to note that the nonlinear model is only normative when the discounting parameter is tuned exactly, and this  is why we  used this specific phrasing. To clarify this, we have added the word `nonlinear' in parentheses when mentioning the normative model in the prior sentence. We also added the following  footnote: \textcolor{blue}{The `nonlinear' model here refers to the family of models obtained by tuning the discounting rate away from the value defining the normative model. This detuning results in a model that is not normative.} \\[0.5ex]
We describe the 0/1-loss function now in the sentence following its introduction: \textcolor{blue}{The 0/1-loss function gives a one unit penalty on trials in which the decision predicted by the model and the data disagree, and no penalty when they agree. Therefore, minimizing this loss function leads to models that best match the trial-to-trial responses in the data rather than the response accuracy.
} \\

\noindent
{\em Page 4, last paragraph: You need the word `alone': the phrase `kappa does not predict the response accuracy' should probably be `kappa alone does not predict the response accuracy.'} \\[1.5ex]
Thanks for pointing this out. We have modified the text as suggested. \\

\thispagestyle{plain}
\noindent
{\em Figure 1, caption: Might be helpful to title panel C as `h=1' and remind readers in the caption that $hT$ and $SNR$ jointly determine accuracy? And for panel D, perhaps write `Maximal accuracy of the ideal observer, at $T\gg1$?} \\[1.5ex]
Good suggestion. We have added the title to Fig.~1C and the sentence, \textcolor{blue}{``Note that $hT$ and ${\mc S}/\sqrt{h}$ jointly determine accuracy.''} to the caption, and also the qualifier, \textcolor{blue}{at $T\gg1$}, to the caption of Fig.~1D. \\ 

\noindent
{\em Equation 5: Would be helpful to explicitly say here that F represents the SNR}\\[1.5ex]
 We have added the phrase  \textcolor{blue}{``representing the SNR''} to the sentence preceding equation 5. \\

\noindent
{\em Paragraph immediately after equation 5: as written, it sounds a bit like F depending only on those 2 parameters follows from keeping $hT$ fixed, although that is not what you mean.} \\[1.5ex]
We have changed the confusing sentence to: \textcolor{blue}{As indicated, $F(hT, {\mc S}/\sqrt{h})$ only depends on $hT$ and ${\mc S}/ \sqrt{h}$.}  \\

\noindent
{\em Paragraph at the end of Section 3: `To increase the accuracy of an ideal observer, it is not enough to increase both click rates, for instance.' I didn't understand that? If everything else is kept fixed, but both lambdas grow, doesn't SNR grow and hT stay fixed?} \\[1.5ex]
In this situation SNR does not necessarily grow. Indeed, this depends on how the parameters are incremented. For instance, if the lambdas grow along the parabolas shown in Fig.~1D, then SNR stays constant. We have edited the sentence referenced above to explicitly state this. It now reads:
\textcolor{blue}{To increase the accuracy of an ideal observer, it is not sufficient to increase both click rates, for instance, since the SNR stays constant if $\lambda_{\rm high}$ and $\lambda_{\rm low}$ follow  the parabolas shown in Fig.~1D.} \\


\noindent
{\em Figure 3C, label on vertical axis: why is this `relative' accuracy?} \\[1.5ex]
The adjective ``relative'', was meant to highlight the fact that we computed the curvature of the graphs in Fig.~3B, representing functions of the relative error rather than the actual values of $\tilde{h}$ and $\tilde{\gamma}$. We agree that this was confusing, and have removed the word ``relative'' from the plot and added an explanation to the caption, which now says:
\textcolor{blue}{Since the functions in panel \textbf{B} do not depend on the actual values of $\tilde{h}$ and $\tilde{\gamma}$, but rather the relative distance of these parameters from reference values, what we show in this plot are \emph{relative} curvatures. We compare relative curvatures as $\tilde{h}$ and $\tilde{\gamma}$ do not have the same units.} \\

\noindent
{\em The authors state that for Figure 4E, NL-L and L-NL do not converge to zero, whereas NL-NL and L-L converge to zero eventually. This is not very clear in the figure, and it would help the reader if the number of trials was larger to show this more clearly. This panel shows that for a given number of trials (in the figure, $<500$ trials), the fits for the linear model were better than the normative model. How sufficient should the number of trials be for the fits to the normative model to be better than the fits to the linear model? In principle, when the normative model is fit to dataset generated by the normative model, it should eventually have lower relative error than the linear model.} \\[1.5ex]
We ran an additional 84 fitting procedures per model pair for training sets of 1,000 trials. We did not run more simulations as our Monte Carlo sampling method is costly (these 84 simulations took 24 hours to run on a modern 4-core laptop). We report the resulting relative errors in the new Fig.~4E. This figure now shows that the NL-NL curve does fall below the L-NL curve on average after 1000 trials. We also added the following explanation to appendix F of the revised manuscript:
\textcolor{blue}{In Fig~4E, up to trial number 500 on the x-axis, the same fits as in panels C-D were used to compute the relative error (y-axis). Because of the high computational cost of our fitting algorithm (Monte Carlo sampling described above), the points for 1000 trials on the x-axis were computed with only 84 independent fits per model pair (as opposed to 500 for the other points of the figure).} \\

\thispagestyle{plain}
\noindent
{\em Figure 5 needs labels C and D. What was the grid of values used for h and in generating Fig. 5? There are comparisons between 0.1 and 2, but do we see a more deviating trend as a function of noise level? That is, what is the deviation like when noise is 0.5 or 1, for example? It would be great to have a summary plot with a metric of deviation separately for both the linear and normative models. What do we see when the metric is shown as a function of noise? Having this summary figure will help the authors establish the point that `the parameters that minimize expected 0/1-loss are biased, and this bias increases with sensory noise,' and that `the NL model exhibits this bias much more strongly than the L model.'} \\[1.5ex]
We have added the labels C and D to the appropriate panels in Figure 5. We have also specified the grid values used in the simulations in appendix F, using the following added passage:
\textcolor{blue}{For the linear model, we used values of $\gamma_1, \gamma_2$ between 0 and 10, with increments of 0.1. For the nonlinear model, we used values of $h_1, h_2$ between 0 and 2.5, with increments of 0.1.}\\[0.5ex]
Regarding the second suggestion of measuring the bias in parameter recovery for intermediate values of noise, we have added the figure that appears below in this response letter, and is now contained in Appendix G to our manuscript. We reference this figure in the following text we have added to Section 7:
\textcolor{blue}{See Appendix~G for a possible metric of the reported bias, and its dependence on sensory noise for each model class (Fig.~6).} \\


\noindent
{\em Appendix B that derives the SNR has typos. In the sentence right after Eq. (16), the statement is $E[N]=(\lambda_{\rm high}-\lambda_{\rm low})/\sqrt(h)$, as T approaches infinity. Should this be $E[N]=(\lambda_{\rm high}-\lambda_{\rm low})/h$?} \\[1.5ex]
Yes, we have checked  Appendix B thoroughly, and corrected this typo. \\ 

\noindent
{\em Eq. (17) seems to contain an error that multiplies 2 in front of hT*exp(-hT).
In the sentence right after Eq. (17), SNR=E[Delta N]/Var[Delta N]. Should this be SNR=E[Delta N]/sqrt(Var[Delta N])? In the equation that follows this sentence, h is not multiplied in the denominator.} \\[1.5ex]
You're right -- we have now corrected these typos, and also spotted a missing factor of 2 in the denominator of the full SNR expression. The equations should all be correct now.  \\

\noindent
{\em For convenience, one could state what S is also in the Appendix.} \\[0.5ex]
We have added Eq.~(19) in Appendix B as a reminder of Eq.~(4) for the reader. \\

\noindent
{\em Appendix F on the details of the simulations could be more detailed. The authors state that when fitting the nonlinear model, Monte Carlos sampling was used. Please explain further on this fitting method to the extent that the reader can replicate.} \\[1.5ex]
We have added the following passage to Appendix F:
\textcolor{blue}{More specifically, the distribution of the decision variable at decision time for a given clicks stimulus, $p(y_T | \mathcal{T})$, was estimated by simulating 800 independent trajectories. Thus, each trajectory had its own independent realization of sensory noise but the realization of the stimulus (timing of the clicks) was frozen. Once the density of $y_T$ was estimated, the likelihood term, $P(d_k|\mathcal{T}_k,\theta)$ in Eq.~(27), could be estimated. More details on this method, such as how the number of 800 particles was chosen and how this method was validated on the linear model for which the analytical solution is available, may be found in section 3.5.5 of~Radillo (2018).} \\

\noindent
{\em Although the confidence intervals do not seem to contain the true parameter value, are the mismatches statistically significant in Figure 4CD? That is, are the parameter values significantly greater than the true parameter value?} \\[1.5ex]
We would like to note that the whiskers of each box in Fig.~4C,D do not represent confidence intervals. Instead, they represent the interval of values that are not considered outliers. More specifically, if $q_1, q_3$ are the first and third quartiles respectively, then the whiskers define the interval: $[q_1 - 1.5\times (q_3-q_1), \ q_3 + 1.5\times (q_3-q_1)]$. \\[0.5ex]
Instead of testing the hypothesis that the MAP estimates are different than the $\theta_\text{true}$ value, we provide a summary statistic for the training datasets of size 500 trials.
We have added the following sentences to the end of section 6:
\textcolor{blue}{For reference datasets of size 500, 98\% of the 500 MAP estimates in the L-NL fits 
	lie strictly above $\gamma_\text{true}$, versus 50.4\% for the corresponding L-L fits. Similarly, 86.6\% of the estimates in the NL-L fits 
	lie strictly below $h_\text{true}$, versus 44.2\% for the corresponding NL-NL fits.}
We believe this is more informative than a p-value. \\

\noindent
\hrulefill

\noindent
{\bf Publication Decision 2 from Neurons, Behavior, Data analysis, and Theory} on August 26, 2019 \\[1ex]
\noindent
{\bf Editorial board's determination:} Accept \\[1ex]
\noindent
{\bf Comments from the editor.} Thanks a lot for your patience with the first round of reviews. I have now checked your revision and have no further comments, which means the manuscript will be transferred to a ``provisional acceptance" stage.

\thispagestyle{plain}

\end{document}